\documentclass[useAMS,usenatbib]{mn2e}

\usepackage{graphicx}

\usepackage{subfig}

\usepackage{amssymb}
\usepackage{amsmath}

\usepackage{fixltx2e}

\usepackage{times}
\usepackage{verbatim}

\usepackage[hidelinks]{hyperref}

\usepackage{aas_macros}
\usepackage{my_macros}

\graphicspath{{.}}

\title[The escape fraction in the epoch of reionization]{The First Billion Years Project: The escape fraction of ionizing photons in the epoch of reionization}
\author[J.-P. Paardekooper et al.]{Jan-Pieter Paardekooper$^{1,2}$\thanks{E-mail:
paardekooper@uni-heidelberg.de}, Sadegh Khochfar$^{3}$ and Claudio Dalla Vecchia$^{4,5}$\\
$^{1}$Universit\"{a}t Heidelberg, Zentrum f\"{u}r Astronomie, Institut f\"{u}r Theoretische Astrophysik, Albert-–Ueberle-–Str. 2, 69120 Heidelberg, Germany\\
$^{2}$Max Planck Institute for extraterrestrial Physics, PO Box 1312, Giessenbachstr., 85741 Garching, Germany\\
$^{3}$Institute for Astronomy, University of Edinburgh, Royal Observatory, Blackford Hill, Edinburgh EH9 3HJ, UK\\
$^{4}$Instituto de Astrofs\'{i}ca de Canarias, C/ V\'{i}a L\'{a}ctea s/n,38205 La Laguna, Tenerife, Spain\\
$^{5}$Departamento de Astrofs\'{i}ca, Universidad de La Laguna, Av. del Astrof\'{i}sico Franciso S\'{a}nchez s/n, 38206 La Laguna, Tenerife, Spain}

\begin{document}

\date{Accepted ***. Received ***; in original form ***}

\pagerange{\pageref{firstpage}--\pageref{lastpage}} \pubyear{2014}

\maketitle

\label{firstpage}

\begin{abstract}
Proto-galaxies forming in low-mass dark matter haloes are thought to provide the majority of ionizing photons needed to reionize the Universe, due to their high escape fractions of ionizing photons. We study how the escape fraction in high-redshift galaxies relates to the physical properties of the halo in which the galaxies form, by computing escape fractions in more than $75000$ haloes between redshifts $27$ and $6$ that were extracted from the {\it First Billion Years} project, high-resolution cosmological hydrodynamics simulations of galaxy formation. We find that the main constraint on the escape fraction is the gas column density in a radius of $10 \, \pc$ around the stellar populations, causing a strong mass dependence of the escape fraction. The lower potential well in haloes with $\Mvir \lesssim 10^8 \, \Msun$ results in low column densities that can be penetrated by radiation from young stars (age $< 5 \, \Myr$). In haloes with $\Mvir \gtrsim 10^8 \, \Msun$ supernova feedback is important, but only $\sim 30 \%$ of the haloes in this mass range have an escape fraction higher than $1 \%$. We find a large range of escape fractions in haloes with similar properties, caused by different distributions of the dense gas in the halo. This makes it very hard to predict the escape fraction on the basis of halo properties and results in a highly anisotropic escape fraction. The strong mass dependence, the large spread and the large anisotropy of the escape fraction may strongly affect the topology of reionization and is something current models of cosmic reionization should strive to take into account.
\end{abstract}


\section{Introduction}


	

Observations are now probing galaxies in the middle of the reionization epoch, when the gas in the inter-galactic medium was transformed from its initially neutral state into a hot, ionized plasma \citep[e.g.][]{2011MNRAS.418.2074M,2012ApJ...758...93F,2014ApJ...793..115B}. Most likely stars in galaxies are responsible for this transformation, although this heavily depends on the fraction of ionizing photons produced by the stars that make it into the inter-galactic medium, the so-called escape fraction $\fEsc$. The escape fraction is a key parameter in studies of the contribution of the observed galaxy population to reionization \citep[e.g.][]{2012ApJ...752L...5B,2013ApJ...768...71R}, semi-analytic modelling of reionization \citep[e.g.][]{Choudhury:2009db,Pritchard:2010br,Santos:2010hr,Mesinger:2011gm,Raskutti:2012wt,2013MNRAS.428L...1M,2012ApJ...747..100S} and numerical simulations of reionization \citep[e.g.][]{Iliev:2006cl,Trac:2007gg,2012MNRAS.423..558C}. A large effort is going into determining the escape fraction observationally. Except for two objects \citep{2011A&A...532A.107L,2013A&A...553A.106L}, in the local Universe no ionizing radiation has been detected directly \citep{Leitherer:1995hd,Deharveng:2001gr}, although some objects show indirect evidence of photon leakage \citep{2011ApJ...730....5H, 2011ApJ...741L..17Z}. The lack of detections may be partly due to selection bias \citep{2013A&A...554A..38B}, but the objects from which radiation is detected have very low escape fractions, $\fEsc < 4\%$. At $z \sim 1$, no objects with leaking ionizing photons have been detected \citep{Siana:2010bo,2010ApJ...720..465B}, but at $z \sim 3$, the highest redshift at which the opacity of the inter-galactic medium for ionizing photons is approximately less than unity, ionizing photons have been detected in $\sim 10 \%$ of the observed objects \citep{2013ApJ...765...47N}. Attempts to constrain the escape fraction with numerical simulations find ranges between $\fEsc < 10 \%$ \citep{Razoumov:2006id,Razoumov:2007kc,Gnedin:2008ib,Paardekooper:2011cz,2013ApJ...775..109K} and $\fEsc > 80 \%$ \citep{2009ApJ...693..984W,Razoumov:2010bh}, with likely a strong mass and redshift dependence \citep{Yajima:2010fb,2014MNRAS.442.2560W}. Due to the opacity of the inter-galactic medium, we need to mostly rely on numerical simulations to learn about the escape fraction during the epoch of reionization. 

The difficulty of constraining the escape fraction both observationally and numerically makes it uncertain whether galaxies are able to reionize the Universe. The observationally determined escape fractions at low redshifts are too low for galaxies to contribute significantly to reionization, which requires an escape fraction of at least 20\% if the luminosity function is integrated down to $M_{\mathrm{UV}}=-13$ to account for the unobserved galaxy population \citep{2013ApJ...768...71R}. However, the samples for which the escapes fractions are measured are small and may be biased. Furthermore it is not at all clear whether the galaxies for which escape fractions are determined at low redshifts are representative for the population of galaxies during the epoch of reionization. It is highly likely that the primary sources of reionization are galaxies that are currently below the detection limit at $z>6$, a regime currently only probed by simulations. In order to learn more about the sources of reionization, we therefore need simulations of a large sample of low-mass galaxies at $z>6$ to better constrain the escape fraction during reionization.

In this work we study the escape fraction in the largest sample of reionization-epoch galaxies to date. Previously we have shown that these proto-galaxies emit enough photons to reionize their local volume \citep{2013MNRAS.429L..94P} (hereafter Paper I) and are therefore likely the main drivers of reionization. We now turn our attention to the reason why these sources are so efficient reionizers, by studying the dependence of the escape fraction on various physical properties of the host halo. Our very large sample of haloes with high spatial resolution allows us to study the scatter of the escape fraction between similar haloes.

\section{Models and methods}

We compute the escape fraction of ionizing radiation by post-processing the output of two high-resolution cosmological hydrodynamical simulations, that are part of the {\it First Billion Years} (FiBY) simulation suite, with detailed radiative transfer simulations. 

\subsection{The First Billion Years simulation}
The FiBY simulation suite (Khochfar et al in prep., Dalla Vecchia et al. in prep.) was run using a modified version of the smoothed-particle hydrodynamics (SPH) code GADGET \citep{Springel:2005cz} that was previously developed for the {\it Overwhelmingly Large Simulations} (OWLS) project \citep{Schaye:2010jl}. In the OWLS-code, gas cooling is computed using tables for line-cooling in photo-ionization equilibrium for 11 elements \citep{2009MNRAS.393...99W}: H, He, C, N, O, Ne, Mg, Si, S, Ca and Fe, which were computed using the CLOUDY v07.02 code \citep{2000RMxAC...9..153F}. For the FiBY simulations, we have added a full non-equilibrium primordial chemistry network \citep{Abel:1997jz,Galli:1998vg,Yoshida:2006ec} including molecular cooling functions for H$_2$ and HD. In this work we focus on two simulations, FIBY\_S and FiBY. The FiBY\_S simulation is the same as was used in Paper I; it has a volume of 4 Mpc (comoving) that was simulated with $2 \times 684^3$ dark matter and gas particles that was run until redshift 6. The FiBY simulation has a volume of 8 Mpc (comoving) with $2 \times 1368^3$ particles that was run until redshift 8.5. The gas particle mass in both simulations is $1253.6 \, \mathrm{M}_{\odot}$. Haloes were identified using the SUBFIND algorithm \citep{2001MNRAS.328..726S,2009MNRAS.399..497D}, and the redshift evolution of the haloes was followed with a merger tree algorithm \citep{2012MNRAS.421.3579N}. The cosmological parameters employed in the FiBY simulation are: $\Omega_{\textnormal{M}} = 0.265$, $\Omega_{\Lambda} = 0.735$, $\Omega_{\textnormal{b}} = 0.0448$, $H_0 = 71 \rm{\,km} \rm{\ s}^{-1} \rm{Mpc}^{-1}$, and  $\sigma_8 = 0.81$, consistent with the recent WMAP results \citep{2009ApJS..180..330K}.

\subsubsection{Star formation} 

We use a pressure law to model star formation \citep{2008MNRAS.383.1210S}, which is designed to yield results consistent with the Schmidt-Kennicutt law \citep{1959ApJ...129..243S,1998ARA&A..36..189K}. The threshold density above which stars form is set to $n=10 \, \mathrm{cm}^{-3}$. This ensures that at $T = 1,000$ K the Jeans mass is resolved by several hundreds of particles. Because of resolution limitations, we assume a polytropic equation of state for gas with densities above the threshold density. The equation of state is normalised to $P_0/k_{\mathrm{B}} = 100 \, \mathrm{cm}^{-3} \, \mathrm{K}$, with $k_{\mathrm{B}}$ Boltzmann's constant, and the effective adiabatic index is $\gamma_{\mathrm{eff}} = 4/3$. There are two types of stars forming in the simulation, depending on the metallicity of the gas. In low-metallicity gas, the stars that form are generally hotter \citep{Schaerer:2002bm} and follow a more top-heavy initial mass function (IMF) \citep{2004ARA&A..42...79B}. For these so-called population (Pop)~III stars we assume a Salpeter IMF \citep{1955ApJ...121..161S} with an upper limit of $M_{\star,\mathrm{upper}} = 500 \, \Msun$ and a lower limit of $M_{\star,\mathrm{lower}} = 21 \, \Msun$. The upper mass limit is chosen to be the maximum mass that is likely to be attained by Pop~III stars \citep{2004NewA....9..353B,Karlsson:2008io}, while the lower limit is chosen to be consistent with data on the chemical abundances in metal-poor stars \citep{2001ApJ...561L..55O,Omukai:2003gw} and with our assumption of a Salpeter slope. Recent cosmological simulations of Pop~III star formation in which radiative feedback was taken into account show that these stars form with masses within this range \citep{2011Sci...334.1250H, 2012MNRAS.422..290S}.
 
With the increase of the metallicity of the star-forming gas, the IMF switches rapidly from the top-heavy Pop~III IMF to a much more bottom-heavy IMF, characteristic of present-day star formation \citep{2001MNRAS.328..969B,2003Natur.422..869S,2005ApJ...626..627O,Dopcke:2012to}. For the metal-enriched Pop~II stars we assume a Chabrier IMF \citep{Chabrier:2003ki}, which has a Salpeter slope at the high-mass end and extends down to sub-solar masses. The critical metallicity at which the transition in the IMF takes place is uncertain, but cosmological simulations suggest that Pop III supernovae quickly boost the local metallicity above the critical metallicity, so the onset of Pop II star formation does not depend sensitively on this choice \citep{2010MNRAS.407.1003M,2011MNRAS.414.1145M}. We choose a critical metallicity of $Z_{\mathrm{crit}}=10^{-4} \, Z_{\odot}$, with $Z_{\odot}=0.02$, a value roughly consistent with both theory \citep{2001MNRAS.328..969B,2002ApJ...571...30S} and the inferred metallicities of metal-poor stars \citep{2007MNRAS.380L..40F,2011Natur.477...67C}.

\subsubsection{Stellar feedback}\label{sec:stellarFeedback}

Massive stars end their lives exploding as supernova, releasing copious amounts of energy and enriching their surroundings with metals. We model the mechanical supernova feedback by injecting thermal energy in the neighbouring particles  \citep{2012MNRAS.426..140D}. For each Pop~II supernova $10^{51}$ erg is injected once a star particle has reached an age of $30 \, \Myr$, corresponding to the maximum lifetime of stars that end their lives as core-collapse supernovae. For Pop~III supernovae we differentiate between type II supernovae which occur for stellar masses between $21 \, \Msun \lesssim m_{\star} \lesssim 100 \, \Msun$ and pair-instability supernovae which occur for $140 \, \Msun \lesssim m_{\star} \lesssim 260 \, \Msun$. We inject $10^{51}$ erg per supernova for the type II SNe after $10 \, \Myr$ and $3 \times 10^{52}$ ergs after $3 \, \Myr$ for the pair-instability supernovae \citep{2002ApJ...567..532H}. For more details on the thermal feedback implementation and extensive tests we refer the interested reader to \citet{2012MNRAS.426..140D}.
 
Stars enrich their surroundings with metals because heavy elements created in their interiors are released in stellar winds and SN explosions. In our simulations stars continuously release hydrogen, helium and metals; we track the enrichment by all 11 species listed above individually. For Pop~II stars the abundances are calculated from yields for asymptotic giant branch stars and type Ia and type II SNe \citep{2007MNRAS.382.1050T,Wiersma:2009ew}. For Pop~III stars we use different yields for type II SNe \citep{2010ApJ...724..341H} and PISNe \citep{2002ApJ...567..532H}. The mixing of the metals with the surrounding medium is modelled by transferring them to neighbouring SPH particles in proportions weighted by the SPH kernel.

\subsubsection{Reionization by the uniform UV-background}

With the appearance of the first luminous sources, the epoch of reionization begins, and the gas in the inter-galactic medium is transformed into a hot and ionized state. When modelling the formation of the first stars and galaxies it is important to take the effect of reionization into account, because it has profound impact on which haloes are able to actively form stars and accrete gas from the inter-galactic medium.

In the FiBY simulations we do not follow the ionizing radiation from the star particles self-consistently due to the computational challenges of radiation-hydrodynamics simulations at these large scales. Instead reionization is modelled using the uniform UV-background model of \citet{2001cghr.confE..64H}, within the bounds of reionization set by WMAP \citep{2011ApJS..192...18K} and Planck \citep{Collaboration:2013uv}. Prior to $z=12$ we assume that reionization hasn't started yet and compute the ionization state of the gas from the collisional ionization equilibrium rates. Between $z=12$ and $z=9$ we gradually change from collisional ionization equilibrium rates to photo-ionization equilibrium rates to mimic the process of reionization. From $z=9$ onwards we assume that the inter-galactic medium is fully ionized, so the photo-ionization rates are used. Self-shielding of dense gas is incorporated at a density threshold of $n_{\textnormal{shield}} \ge 0.01 \, \cmmt$, above which the flux is decreased with a fraction $(n/n_{\textnormal{shield}})^{-2}$ \citep{2010ApJ...725L.219N,2012MNRAS.427.2889Y}. This fraction is proportional to the recombination rate, so there is a continuous transition between shielded and unshielded regimes. The rates in the shielded gas are computed by interpolating between the collisional and photoionization equilibrium tables.

The disadvantage of using a uniform UV-background is that reionization happens independently from the actual number of ionizing photons produced by the stars that have formed in the computational volume. In the FiBY simulations the time-scale in which the UV-background starts to take effect is consistent with the ionizing photon output from proto-galaxies (Paper I, but see \autoref{sec:radiative_transfer}). The disadvantage of assuming that reionization takes place uniformly is that we neglect the spatial distribution of the reionization sources. The effects of reionization will therefore happen too early in some regions and too late in other regions. On the other hand, this approach has the advantage that reionization affects all haloes at the same time. This way we can study the effect of reionization on the escape fraction in the entire halo sample, and disentangle the effects of external feedback by the UV-background and internal feedback by supernova explosions.

\subsection{Computation of the escape fraction of ionizing photons from proto-galaxies}

The FiBY\_S and FiBY simulations that we focus on in this work do not contain the rare high density peaks where the most massive galaxies form due to the limited volume of these simulations. However, high resolution is essential for the computation of the escape fraction, since the main constraint on the escape fraction is the local environment of the sources \citep{Paardekooper:2011cz}. Furthermore, this high resolution is necessary to resolve star formation in the abundant low-mass proto-galaxies that are likely the main drivers of reionization \citep[Paper I,][]{2014MNRAS.442.2560W}.

From these simulations we have extracted all the dark-matter haloes that contain at least 1000 dark-matter particles, 100 gas particles and 1 star. We chose a lower limit of 1000 dark-matter particles to make sure that there are enough particles to represent the halo properties and minimise resolution effects in the lowest mass regime. The lower limit on the halo mass coincides with the mass in which the first stars are forming, so this choice has small impact. Similarly, we chose a lower limit of 100 gas particles to avoid bias in the radiative transfer simulations. The particle-to-particle transport scheme used in SimpleX becomes inaccurate if there are not enough gas particles to represent the gas in the halo. This lower limit may lead us to exclude the haloes in which supernova feedback cleared away all the gas, but we were careful to choose the limits in such a way that there are only a few haloes excluded at redshifts higher than $12$. At lower redshifts there are low-mass haloes that we exclude because the gas has been cleared away by feedback from the UV-background. Since there is no active star formation going on in these haloes, excluding these haloes will not affect our results. The number of haloes we obtained for post-processing with radiative transfer is 11364 for the FiBY\_S simulation and 64319 for the FiBY simulation, resulting in a total of 75801 haloes between redshift $27$ and $6$, in the mass range $2 \times 10^6 - 6 \times 10^9 \, \Msun$, several orders of magnitude more than in previous studies of this kind \citep{2009ApJ...693..984W,Razoumov:2010bh,2014MNRAS.442.2560W}.

In every halo we compute the escape fraction by comparing the number of photons that are produced by the stars  $N_{\mathrm{emitted}}$ to the number of photons that reach $r_{200}$ of the main subhalo, $N_{\mathrm{phot}}(r > r_{200})$:
\begin{equation}\label{eq:fEsc}
  f_{\mathrm{esc}} = \frac{N_{\mathrm{phot}}(r \ge r_{200}) }{N_{\mathrm{emitted}}},
\end{equation}
where $r_{200}$ is the radius at which the overdensity is 200 times the critical density. The number of photons that escape from the halo are computed using detailed radiative transfer simulations, the details of which we describe below.

\subsubsection{Radiative transfer}\label{sec:radiative_transfer}

The radiative transfer simulations were carried out with a modified version of the SimpleX algorithm \citep{Paardekooper:2010iu}. SimpleX computes the transport of ionizing photons on an unstructured grid created from the positions of the SPH particles. The particles are connected to their nearest neighbours by means of the Delaunay triangulation \citep{Delaunay:1934um} and grid cells are formed by the Voronoi tessellation \citep{Dirichlet:1850ut,Voronoi:1908ux}, which is the mathematical dual of the Delaunay triangulation. Contrary to the SPH kernel, this results in a consistent volume discretisation \citep{Hess:2010if}.
 
Ionising photons are transported from particle to particle using the connectivity of the Delaunay grid. At every grid point interaction of the photons with the gas takes place, in which case we compute the absorption of photons by neutral atoms, taking into account the path length through the cell, and recombinations of ionized atoms. This method of transporting photons has the advantage that the computation time does not scale linearly with the number of sources in the simulation. This is a necessary requirement for the computation of the escape fraction, as the most massive haloes contain more than 140,000 star particles. Reducing the number of sources by merging star particles or using only the youngest sources that contribute the majority of ionizing photons could lead to incorrect results \citep{Paardekooper:2011cz}.
 
Among the modifications to the method is the self-consistent computation of the temperature of the gas as described in \citet{2010PhDT........63P}, including photo-heating, recombination cooling, collisional ionization cooling, collisional excitation cooling and cooling from free-free emission. To compute the photo-heating accurately, the spectrum of the sources is taken into account by dividing the photon packets in multiple frequency bins. 

\begin{figure} 
  \includegraphics[width=85mm]{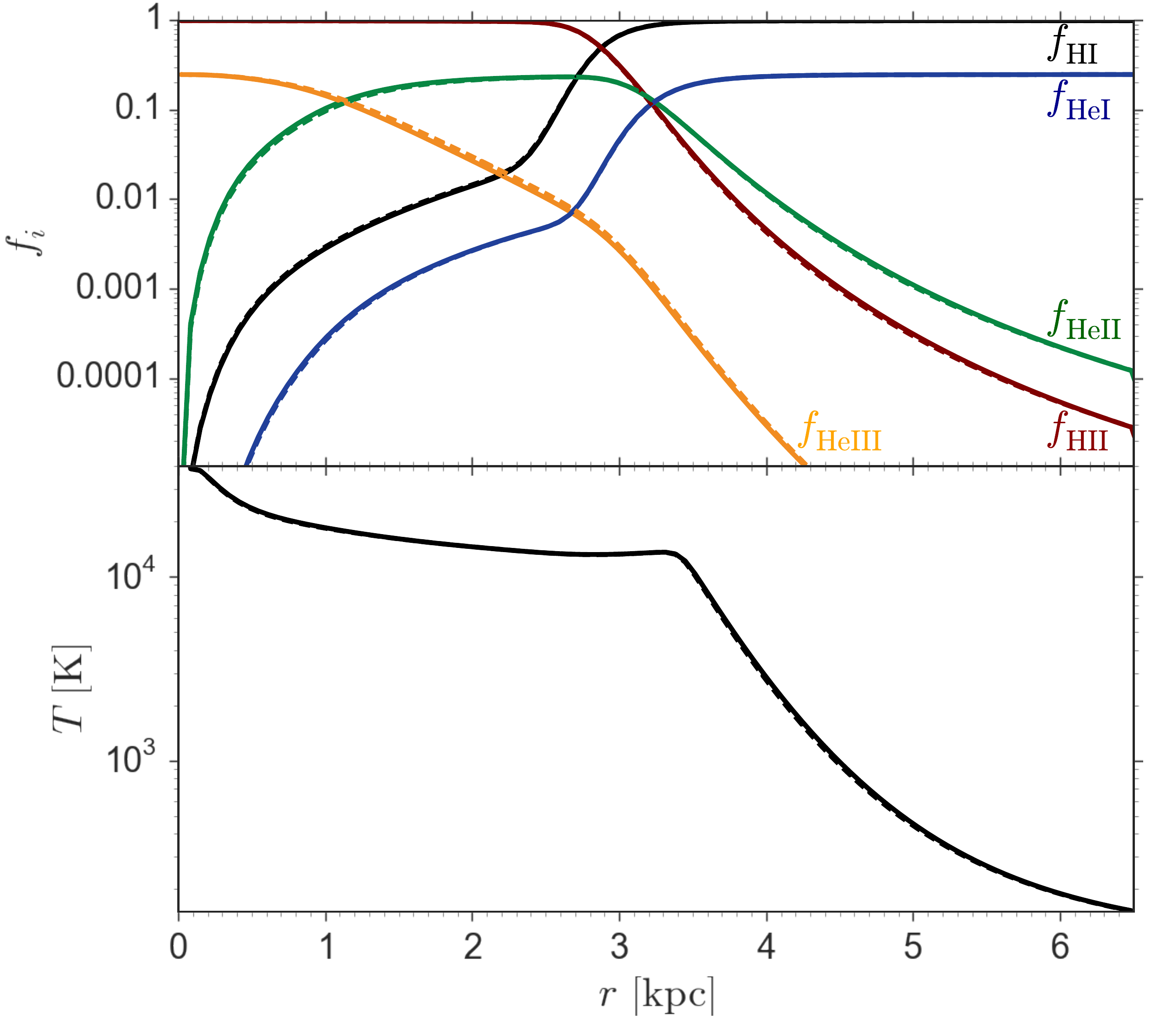}
  \caption{ 
Results of a 1-dimensional simulation of the ionized region around a single source including hydrogen and helium. The simulation contains $100$ particles and was run for 30 Myr. The computational domain is $6.6$ kpc. A single source is located at the edge and is assumed to have a spectrum of a $100,000$K black body. Solid lines depict a simulation with 100 frequency bins evenly spaced in logspace, while the dashed lines depict a simulation with 10 frequency bins spaced to optimally represent the source spectrum and cross sections.  {\it Upper panel:} Spherically averaged species fraction as a function of radial distance from the source. {\it Lower panel:} Spherically averaged temperature as a function of radial distance from the source. 
  }
  \label{fig:Test2}
\end{figure}

We have furthermore included the absorption of ionizing photons by helium atoms and the corresponding heating and cooling processes in the temperature computation. We have tested the new implementation in standard test problems. In \autoref{fig:Test2} we show the results of the new solver in a 1-dimensional simulation with $100$ particles and $1$ source sampled with 10 frequency bins. The size of the computational domain is $6.6$ kpc, with the source located at the edge of the domain. The source emits $5 \times 10^{48}$ ionizing photons per second with a spectral shape of a black body of temperature $100,000$ K. The density of the gas is $0.001 \, \mathrm{cm}^{-3}$; we assume the gas to be of primordial composition with a hydrogen mass fraction of $0.75$ and a helium mass fraction of $0.25$. The gas is initially neutral, with a temperature of $100$ K. SimpleX is capable of transporting diffuse recombination radiation, however, for easy comparison to the results of other codes we assume the on-the-spot approximation\footnote[1]{In the on-the-spot approximation it is assumed that all ionizing photons produced by recombinations directly to the ground state are immediately reabsorbed.} in this test. We show in \autoref{fig:Test2} the temperature of the gas and the species fractions after $30$ Myr. We find that these are in excellent agreement with the results obtained with other state-of-the-art radiative transfer codes \citep{2009MNRAS.393..171M,2011MNRAS.411.1678C,Pawlik:2011ei,2012MNRAS.421.2232F}.

\begin{figure} 
  \includegraphics[width=85mm]{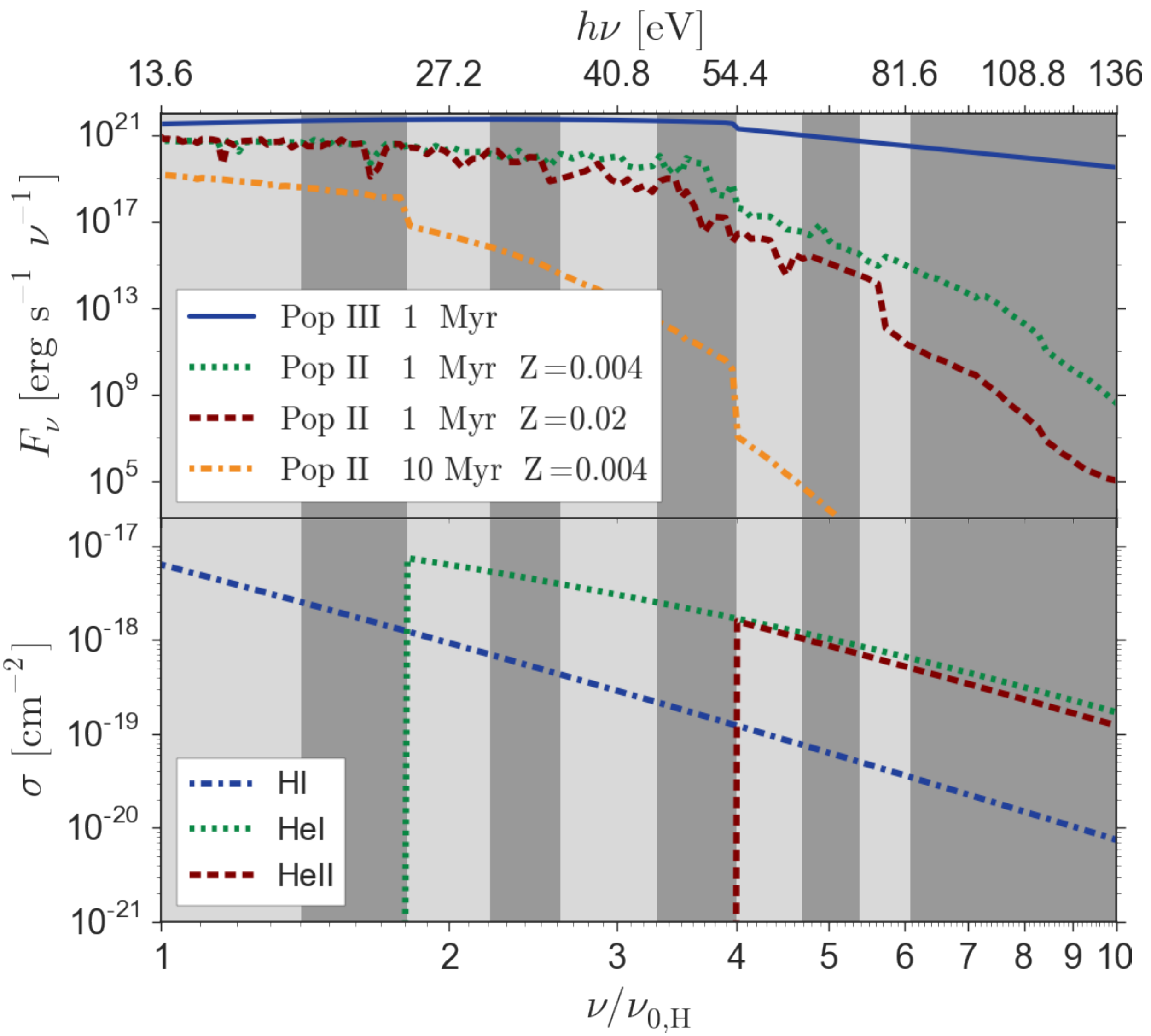}
  \caption{ 
Frequency bin spacing in the radiative transfer simulations. {\it Top panel:} Fluxes for four different source types as a function of frequency normalised to the hydrogen ionization frequency, $\nu_{0,H}=3.3 \times 10^{15} \, \mathrm{Hz}$. The blue line denotes the flux of a Pop~III star of $1$ Myr, the green line the flux of a Pop~II star of $1$ Myr with metallicity $Z=0.004$, the red line the flux of a Pop~II star of $1$ Myr with metallicity $Z=0.02$ and the orange line the flux of a Pop~II star of $10$ Myr with metallicity $Z=0.004$. The shaded areas represent the frequency bin spacing. {\it Bottom panel:} The cross sections of neutral hydrogen, neutral helium and singly ionized helium as a function of frequency. The shaded areas represent the frequency bin spacing.}
  \label{fig:freqBins}
\end{figure}

For the FiBY project we have added a new prescription for the sources in the simulation, so that we no longer have to assume that the source spectra follow a black-body spectrum as in \citet{Paardekooper:2010iu} and \citet{Paardekooper:2011cz}. The ionizing emissivity and spectrum of the star particles are computed from stellar synthesis models for both Pop~III \citep{Raiter:2010hs} and Pop~II \citep{Leitherer:1999jt} stars, taking into account the age and metallicity of the stellar population. There is considerable uncertainty in the total number of ionizing photons produced by a Pop~II stellar population between different stellar synthesis models, depending on the physics included. For example, the BPASS \citep{Eldridge:2009tw} models include a more detailed treatment of massive stellar binaries which leads to a factor $\sim 1.5$ higher ratio of ionizing to non-ionizing UV flux compared to the GALAXYEV \citep{2003MNRAS.344.1000B} models \citep{Mostardi:2013ff}. Adding the effect of rotation could even lead to a factor of $\sim 5$ difference in the ionizing photon production \citep{2014ApJS..212...14L}. A change in the number of ionizing photons produced by the stars has a large effect on the escape fraction, so it is important to keep these uncertainties in mind.

Compared to the simulations presented in Paper I we have made one important change. Whereas in Paper I we computed the spectra for the Pop~II stars using the GALAXYEV package \citep{2003MNRAS.344.1000B}, for consistency with subsequent studies of the FiBY simulations we here use the Starburst99 spectra \citep{Leitherer:1999jt} for the radiative transfer simulations. We compare the total ionizing emissivity of the haloes between Starburst99 and GALAXYEV in \autoref{sec:appendix_s99vsbc03}, showing that during the epoch of reionization both models provide similar results in terms of ionizing emissivity, and only deviate from each other at $z < 7$, when old populations contribute more ionizing photons in \citet{2003MNRAS.344.1000B} than in Starburst99.

In Paper I we have computed the number of ionizing photons produced by the stars assuming continuous star formation in the star particles, which was necessary for the interpolation between snapshots in the reionization model, and to account for ionization of the gas by previous episodes of star formation that we miss due to the post-processing nature of our simulations (see \citet{Razoumov:2006id} for a similar approach). We have found that due to the fast recombination time-scale of the gas inside the haloes, previous episodes of star formation have no significant impact on the escape fraction. The assumption of constant star formation in Paper I is therefore a strict upper limit on the ionizing emissivity, especially in low-mass haloes where only a single stellar population formed. Although taking into account the effect of photo-heating on the gas would increase the recombination time (both by increasing the temperature and decreasing the density through D-type ionization fronts), we have decided to take a more conservative approach in this work and assume a burst mode for star formation. The difference in escape fractions between Paper I and this work show the limits of the post-processing approach for the computation of escape fractions. In \autoref{sec:appendix_constVsBurstSF} we show the difference in escape fraction between the two approaches. Unfortunately it is extremely computationally demanding to perform cosmological radiation hydrodynamics simulations with the resolution and the level of detail in the radiative transfer that we present here, so in this work we have to rely on post-processing.

In the simulations presented here, the source spectrum is sampled in 10 frequency bins, which is sufficient to eliminate errors associated with finite frequency resolution in simple test cases using black-body and power-law sources \citep{Mirocha:2012wf}. In \autoref{fig:Test2} the solid lines depict a simulation with 100 frequency bins evenly spaced in log-space. We have verified that the simulation has converged for this number of frequency bins. The dashed lines depict a simulation with 10 frequency bins spaced to optimally represent the spectrum and cross sections, showing no difference with the converged simulation. In \autoref{fig:freqBins} we show the frequency bin spacing together with examples of source spectra and the cross sections for neutral hydrogen, neutral helium and singly ionized helium. We have verified that this frequency bin spacing in combination with the source spectra used in this work gives similar accurate results as the black-body case shown in \autoref{fig:Test2}.

\subsubsection{Creating all-sky maps}\label{sec:ray_trace}

SimpleX computes the ionization state of the gas and the escape fraction of ionizing photons on the basis of a Delaunay triangulation of the SPH particles. This means that the angular resolution of the radiative transfer calculation at the virial radius where the escape fraction is computed, is determined by the number of SPH particles there. The size of the Voronoi cells in which the radiative transfer is computed depends on the position of the particles, resulting in different cell sizes at the virial radius. This is no restriction in computing the escape fraction: this depends on the total number of absorptions inside the virial radius, and thus on the gas density which is computed from the SPH particles. However, to determine the angular distribution of the escape fraction we need to know the escape fraction in equal-area pixels on a sphere at the virial radius. For this reason we use a different technique to generate full-sky maps, namely post-processing the output from the radiative transfer simulations with a ray-tracing routine. 

We define a sphere at $r_{200}$ that we pixelate using the HEALPix algorithm \citep{2005ApJ...622..759G} with $3072$ pixels. Rays are cast from every source to every pixel, computing the number of photons along the ray that is not absorbed and reaches $r_{200}$, which is given by
\begin{equation}
    N_{r_{200}}^{\mathrm{ray}} = N_{\mathrm{emitted}}^{\mathrm{ray}} \, e^{-\tau},
\end{equation}      
where $\tau$ is the optical depth along the ray. We then add up the contributions of all rays in every pixel to compute the escape fraction on the full-sky map. We compute $\tau$ by averaging the number density and ionization state of hydrogen and helium in every tetrahedron on the Delaunay grid over the 4 vertices and computing the length of the ray through the tetrahedron by calculating the intersection between the ray and the triangles that constitute the faces of the tetrahedron \citep{Moller:1997tm}. Because the ionization state of the gas is already given by the SimpleX simulations, we only need to cast every ray once. Even so, due to the scaling of the computation time with the number of sources and the number of pixels, for the largest haloes this computation takes significantly longer than the full radiative transfer simulations with SimpleX. 

In most haloes the escape fraction computed with the ray-tracing method is lower than computed with SimpleX. This is caused by the way the optical depth is computed in both methods. In SimpleX ionizing photons travel along the edges of the triangulation so the optical depth along the path is determined by the density and the ionization state of a single particle \citep[see Fig. 5 in][]{Paardekooper:2010iu}. In the ray-tracing routine the optical depth is computed through tetrahedra that consist of 4 particles, the density and ionization state are averaged over these 4 particles. This in general leads to an overestimate of the optical depth in the ray-tracing approach. For our purposes this is not a big issue since the relative escape fraction between pixels is not likely to be profoundly affected by this.

\section{Results}

%

\subsection{Absorption of the ionizing radiation}\label{sec:absorption}

The escape fraction of ionizing radiation in a halo depends solely on the fraction of photons that is absorbed. If we consider a halo of uniform (hydrogen-only) gas density with a single monochromatic source in the centre, whether any ionizing photons will escape depends on whether the Str\"{o}mgren sphere lies within the virial radius or not. In other words, if recombinations are able to balance the ionizations from the source photons, the escape fraction will be zero. In the limit of completely ionized gas the recombination rate has a strong dependence on the gas density, scaling as $n_{\mathrm{H}}^2$.

Although the gas density is generally not uniformly distributed in galaxies, the gas density around the sites of star formation can be so high that the recombination rate of the gas prevents any photons from escaping. In idealised simulations of disc galaxies this prevents any ionizing photons from escaping except when feedback clears away (part of) this dense gas \citep{Paardekooper:2011cz}. In our much larger sample we can now study whether this is also the case for haloes in a cosmological simulation. 

\begin{figure*}
 \includegraphics[width=\textwidth]{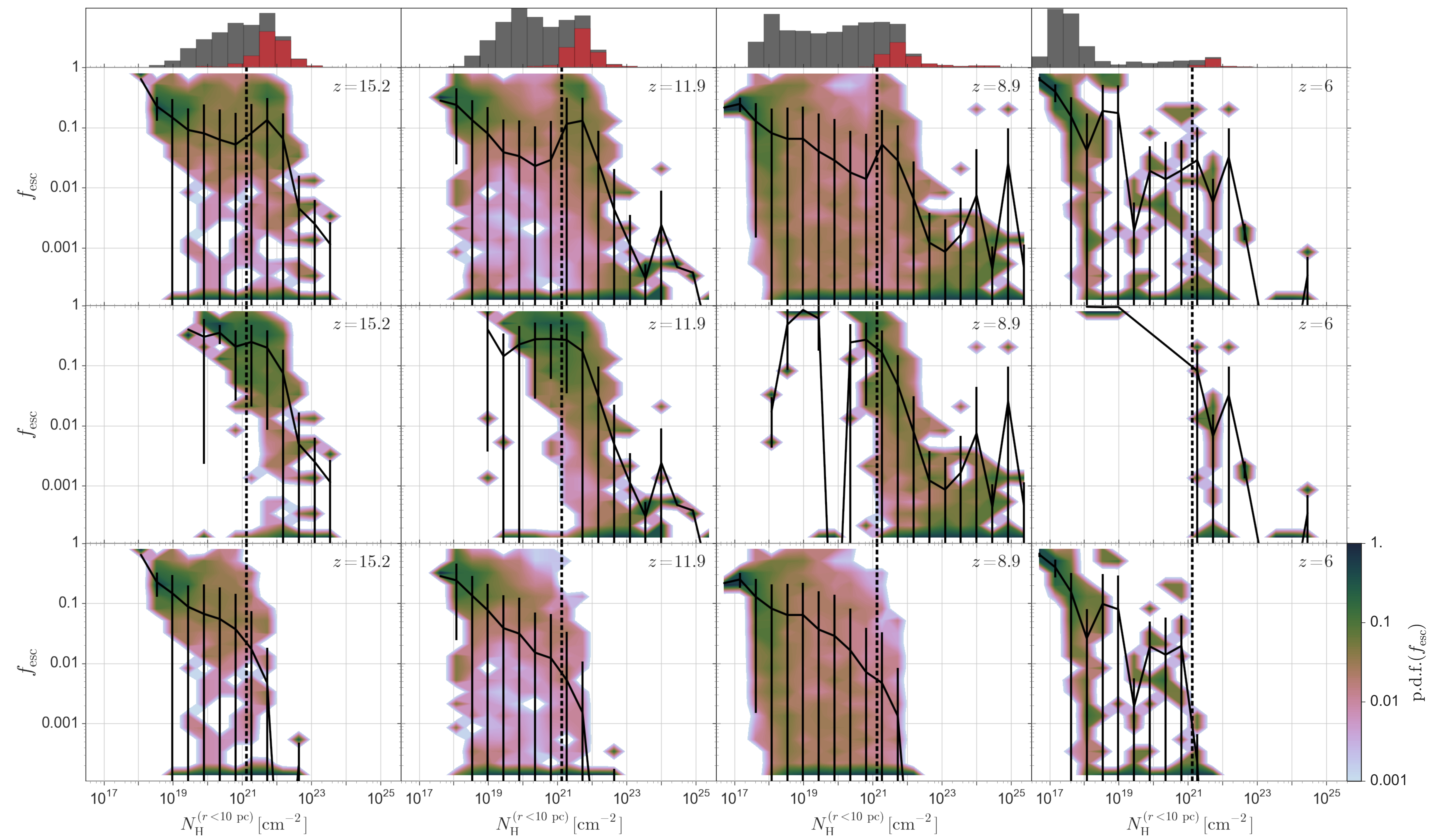}
  \caption{The total escape fraction in a halo as a function of $\NH$ at different redshifts. Here, $\NH$ is the spherically averaged column density within 10 pc of a source, computed as a ionizing-emissivity-weighted average over all sources in a halo. Colour-coded is the 2-dimensional histogram of all the haloes within a mass bin, the solid line represents the mean escape fraction in the same bin. The errors are the standard deviation around the mean. For clarity, haloes with $\fEsc < 10^{-4}$ are plotted at $\fEsc = 10^{-4}$. The top row includes all haloes, the middle row only haloes with at least one star particle younger than $5 \, \Myr$ and the bottom row only haloes containing no star particle younger than $5 \, \Myr$. The dotted vertical line indicates the column density at which the Str\"{o}mgren sphere around a star particle of age $5 \, \Myr$ or younger would be $10 \, \pc$. The histogram on the top shows the number of haloes in each bin (grey) and the number of haloes containing at least one star particle younger than $5 \, \Myr$ (red). At redshift $6$ the data is from the FiBY\_S simulation only, resulting in larger Poisson errors than the higher redshifts.}
  \label{fig:fEsc_NH}
\end{figure*}

In \autoref{fig:fEsc_NH} we show the total escape fraction in a halo as a function of $\NH$, where $\NH$ is the spherically-averaged column density within $10$ physical $\pc$ around a source, taken as the ionizing-emissivity-weighted average over all sources in the halo:
\begin{equation}\label{eq:NH} 
    \NH = \frac{\sum_{i} N_{\textnormal{H},i}^{(r < 10 \, \pc)}  N_{\textnormal{ion},i}}{\sum_{i} N_{\textnormal{ion},i}},
\end{equation}
where the sums are over all star particles in a halo. This quantity gives us an indication of how dense the gas around the sources in the halo is on average. The radius of $10 \, \pc$ was chosen to be close to the minimum smoothing length of the simulation at all redshifts where star formation is taking place. We have verified that increasing the sphere radius does not affect these results, indicating that in most haloes the ionizing photons are indeed absorbed in the immediate vicinity of the sources. This has to do with the density profile of the gas in the haloes: the dense gas is concentrated where the star formation is taking place. However, this picture is likely oversimplified as the high-redshift Universe is a violent environment for proto-galaxies and the frequent minor mergers will cause more disturbed systems. At redshift $6$ the data is from the FiBY\_S simulation only, resulting in larger Poisson errors than the higher redshifts due to smaller statistics.

The top row of \autoref{fig:fEsc_NH} shows the results for all haloes in the sample. At all redshifts $\NH \lesssim 10^{19} \, \cm^{-2}$ means that on average $\fEsc \gtrsim 0.1$. The column density in the vicinity of the sources is so low that star particles of all ages are able to keep the gas ionized, resulting in a high escape fraction. As $\NH$ increases up to $\NH \sim 10^{21} \, \cm^{-2}$, $\fEsc$ decreases, as it is getting more and more difficult for the ionizing photons to penetrate the higher column density. If the dense gas is fully ionized and uniformly distributed in a sphere around the sources, the number of absorptions would scale as $n_{\mathrm{H}}^2$. In that case the decrease of $\fEsc$ would scale as $1 - n_{\mathrm{H}}^2$, which means that at $\NH \gtrsim 10^{19} \, \cm^{-2}$ the escape fraction would be close to zero. However, at all redshifts we find that the scaling of the escape fraction is a much less strong function of the column density. This indicates that the assumption of a uniform distribution of the dense gas around the sources in \autoref{eq:NH} is a simplification and that the spatial distribution of the gas is important.

At $\NH \gtrsim 10^{21} \, \cm^{-2}$ there is an increase in the escape fraction. At these column densities lies the threshold for star formation in the simulation. If we again assume a constant gas density in a sphere with radius of $10 \, \pc$ around the sources, the star formation threshold of $n = 10 \, \cm^{-3}$ translates into $\NH \simeq 3 \times 10^{20} \cm^{-2}$. This is around the column density at which we see the increase in escape fraction. Haloes with $\NH \lesssim 3 \times 10^{20} \cm^{-2}$ contain mostly stellar populations that have cleared away the dense gas through supernova feedback. Since supernova feedback from Pop II stars sets in after $30 \, \Myr$ (see \autoref{sec:stellarFeedback}), these populations are relatively old. The majority of ionizing photons that are produced by a Pop II stellar population over time are emitted in the first $\sim 5 \, \Myr$, because that is when the massive, hot stars are still around (see \autoref{fig:freqBins}). Although stellar populations younger than $5 \, \Myr$ are generally surrounded by column densities $\NH \gtrsim 3 \times 10^{20} \cm^{-2}$, these populations are able to penetrate higher column densities: a Pop II stellar population younger than $5 \, \Myr$ is able to fully ionize a homogeneous cloud of radius 10 pc if the column density is less than $1.3 \times 10^{21} \, \cm^{-2}$. We have indicated this column density with a vertical dashed line in \autoref{fig:fEsc_NH}. As $\NH$ gets higher than this value, $\fEsc$ gets lower, as even the young stellar populations are not able to penetrate these high column densities.

To further clarify this point, we show in the middle and bottom row of \autoref{fig:fEsc_NH} the subset of haloes containing at least one young stellar population and haloes containing no young stellar populations, respectively. The two subsets show a similar relation: a decrease in escape fraction as $\NH$ gets higher, albeit shifted because young stellar populations are able to penetrate much larger column densities. Although the drop in escape fraction for the haloes containing young stars is more steep than for the haloes containing only old populations, the scaling with density is less strong than $1 - n_{\mathrm{H}}^2$. The reason for this is two-fold. In $\sim 30 \%$ of the haloes there are a few young stellar populations which are unobscured by dense gas. Shortly after these populations had formed, a supernova event close by cleared away the gas around these sources. The ionizing photons from these sources can therefore travel through gas with much lower column density, causing a high overall escape fraction. In the majority of haloes ($\sim 70 \%$) the spherically averaged column density around all the young stellar populations is high enough to shield all the produced ionizing radiation. In these cases the non-zero escape fraction is caused by the inhomogeneous distribution of the dense gas around the sources. The resulting escape fraction at the virial radius will therefore be highly anisotropic.

\begin{figure*}
 \includegraphics[width=\textwidth]{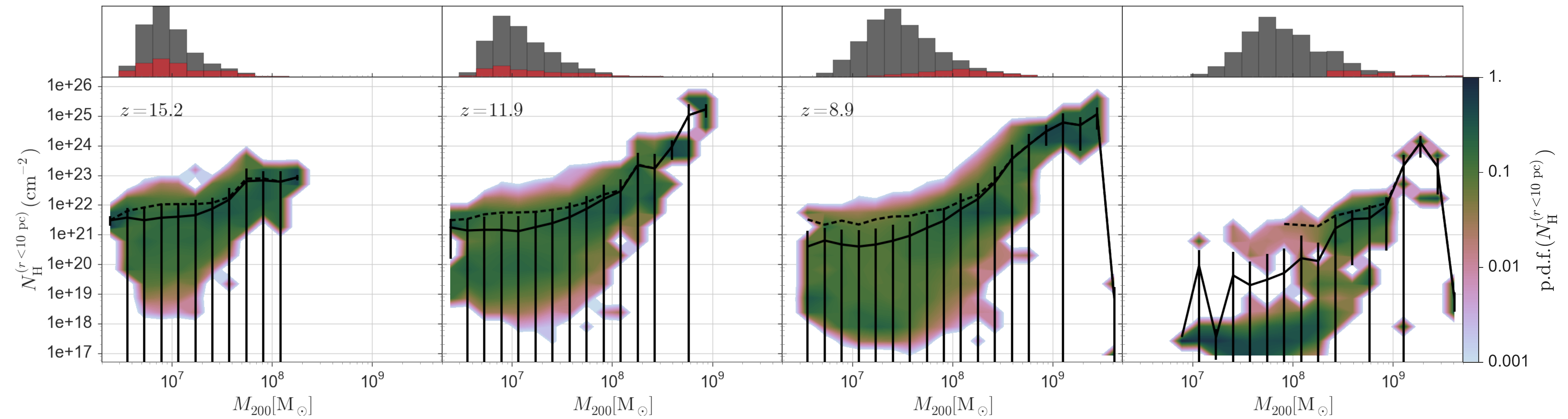}
  \caption{The spherically averaged column density within 10 pc of a source, computed as a ionizing-emissivity-weighted average over all sources in a halo, as a function of halo mass $\Mvir$ for different redshifts. Colour-coded is the 2-dimensional histogram of all the haloes within a bin, the solid line represents the mean column density in the same bin and the dashed line the mean column density for haloes containing at least one stellar particle with age equal to or less than 5 Myr. The errors are the standard deviation in the mean. The histogram on the top shows the number of haloes in each bin (grey) and the number of haloes containing at least one star particle younger than $5 \, \Myr$ (red). At redshift $6$ the data is from the FiBY\_S simulation only, resulting in larger Poisson errors than the higher redshifts.}
  \label{fig:NH_mVir}
\end{figure*}

For the later discussion it is useful to link $\NH$ to other halo properties. In \autoref{fig:NH_mVir} we show the dependence of $\NH$ on the virial mass of the halo, defined as the total mass of the halo (dark matter, gas and stars) within the virial radius. We estimate the virial radius using $r_{200}$, the radius at which the overdensity is 200 times the critical density. A higher halo mass means that the average column density around the sources is also higher, due to the deeper potential well there is more gas in the centre of the halo, where stars are forming. For haloes with $\Mvir \gtrsim 10^{8} \, \Msun$, $\NH$ and $\Mvir$ show a tight relationship. For lower masses the scatter is larger due to the efficient redistribution of the dense gas by supernova feedback in lower-mass haloes. At redshift $6$ the effect of the UV-background on haloes of $\Mvir \lesssim 2 \times 10^8$ is clear: the column density around the (primarily old) sources is low due to photo-heating by the UV-background. Gas in higher-mass haloes is able to self-shield. There are only a few haloes in the highest mass bins, so the drop in $\NH$ at high masses at redshifts $8.9$ and $6$ is likely caused by statistics rather than physics.

\subsection{The dependence of the escape fraction on the halo properties}\label{sec:halo_properties}

In the previous section we have shown that the escape of ionizing radiation from a halo is mainly driven by the absence of dense gas in the vicinity of the most luminous sources. Here we show how the escape fraction scales with different halo properties.

\subsubsection{Mass}

\begin{figure*} 
  \includegraphics[width=\textwidth]{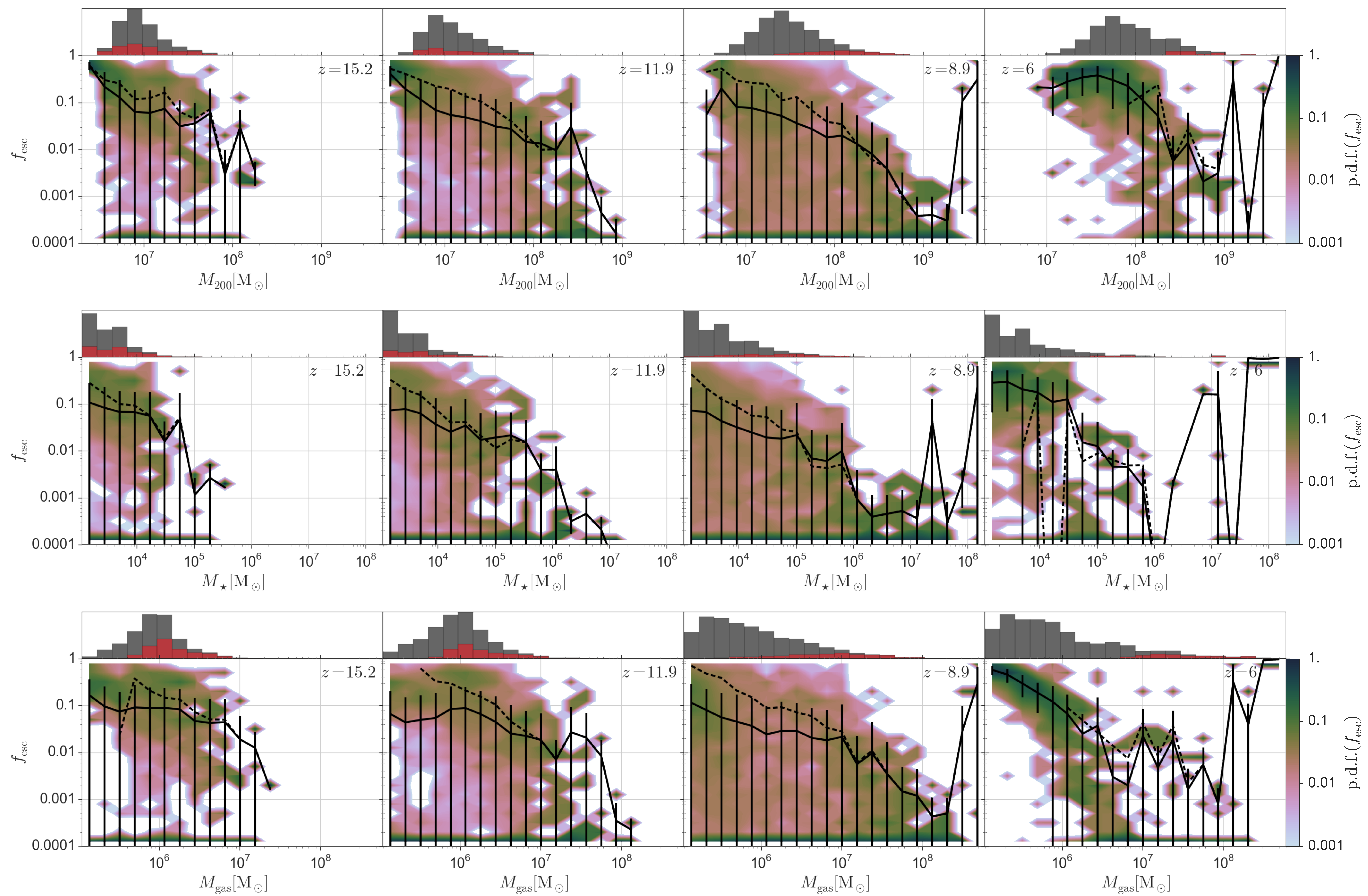}
  \caption{The escape fraction as a function of halo mass $\Mvir$ (top), stellar mass $\Mstar$ (centre) and gas mass $\Mgas$ (bottom) for different redshifts. Colour-coded is the 2-dimensional histogram of all the haloes within a mass bin, the solid line represents the mean escape fraction in the same bin and the dashed line the mean escape fraction for haloes containing at least one stellar particle with age equal to or less than 5 Myr. The errors are the standard deviation in the mean. For clarity, haloes with $\fEsc < 10^{-4}$ are plotted at $\fEsc = 10^{-4}$. The histogram on the top shows the number of haloes in each bin (grey) and the number of haloes containing at least one star particle younger than $5 \, \Myr$ (red). At redshift $6$ the data is from the FiBY\_S simulation only, resulting in larger Poisson errors than the higher redshifts.}
  \label{fig:fEsc_mass}
\end{figure*}

In \autoref{fig:fEsc_mass} we show the escape fraction as function of halo mass, using three different probes of the mass. The first is the virial mass, defined in the previous section. The haloes we consider in this work have virial masses in the range $2 \times 10^{6} \, \Msun \lesssim \Mvir \lesssim 6 \times 10^{9} \, \Msun$. The lowest mass is set by the smallest haloes in which star formation takes place, while the largest mass is a result of the size of the simulation volume. Higher mass haloes than what we find do form at these redshifts, but they form in rare high-sigma peaks that our simulation volume does not capture. However, the mass range we cover is where the majority of ionizing photons during reionization is expected to be produced \citep{Barkana:2001tz}. The stellar mass in the haloes lies between $1250 \, \Msun \lesssim \Mstar \lesssim 2 \times 10^{8} \, \Msun$, where the minimum mass is given by the SPH particle mass. The gas mass in the haloes lies in the range $1 \times 10^{5} \, \Msun \lesssim \Mgas \lesssim 8 \times 10^{8} \, \Msun$. 

We find a strong dependence of the escape fraction on mass, independent of which quantity we use to specify the mass. Mini-haloes with virial mass $\Mvir \lesssim 3 \times 10^{6} \, \Msun$ always have an escape fraction higher than $10 \%$, due to their low gas content. These haloes form primarily Pop III stars which are luminous enough to ionized most of the gas in the halo, hence the high escape fraction. The average escape fraction for haloes with $\Mvir \sim 10^{7} \, \Msun$ is still high ($\gtrsim 10 \%$), but there are many haloes with zero escape fraction. Haloes in this mass range that contain at least one young star particle have high escape fractions, because star particles younger than $5 \, \Myr$ are able to ionize most of the gas in the halo. The high escape fraction is caused by a combination of high ionizing photon production and relatively low column density (see \autoref{fig:fEsc_NH}) due to the small gas mass. The majority of these haloes contain only Pop II stars, so this is not caused by supernova feedback which sets in too late, when the massive stars that produce the majority of ionizing photons have already disappeared. However, in haloes in this mass range without young stars the column density of the gas close to the sources is high enough to absorb all the ionizing radiation from older star particles, so for radiation to escape supernova feedback must clear away part of the dense gas surrounding the sources in order for the escape fraction to be higher than zero. As we move to higher halo masses, the average escape fraction gets lower. Due to the higher column densities in this mass range, most of the ionizing radiation produced by young star particles is absorbed.  With a few exceptions, the escape fraction in haloes with mass $\Mvir \gtrsim 10^{9} \, \Msun$ is below $0.1 \%$. We find the similar behaviour of the escape fraction as a function of stellar mass and gas mass. In the highest mass bins at redshift $8.9$ and $6$ the average escape fraction is higher than $10 \%$. This is caused by recent feedback events in some of the massive haloes that have cleared away the dense gas surrounding the sources (see \autoref{fig:NH_mVir}). Unfortunately there are not enough haloes in this mass range to distinguish whether this is a physical effect or whether it is caused by the statistics.

In the previous section we found a clear distinction between the escape fraction in haloes containing at least one star particle with age equal to or below 5 Myr and the haloes containing only older populations. Due to the higher production of ionizing photons in young stellar populations, for a given column density around the sources, the total escape fraction is higher in the haloes that are actively forming stars. In \autoref{fig:fEsc_mass} the dashed line represents the mean escape fraction for haloes containing young stars. While at the high-mass end there is no difference (these haloes have enough mass to continuously form stars), haloes with virial mass $\Mvir \lesssim 10^{8} \, \Msun$ that contain a young stellar population have on average a higher escape fraction. The reason for this is that there are less haloes with zero escape fraction compared to the total sample, because the low-mass haloes do not contain enough dense gas to shield the radiation from the luminous young stars.

The relation between the escape fraction and the mass is similar for the redshift range $15 \gtrsim z \gtrsim 9$, for all three mass probes. At lower redshifts there are massive haloes that were not present at higher redshifts, but in the mass-range probed in all three highest redshift bins the average escape fraction does not change much. This is different from the results in Paper I where we found that the escape fraction would be higher at lower redshift in a constant mass bin. This difference is caused by the way the ionizing emissivity is computed: the continuous star formation rate model used in Paper I overestimates the ionizing emissivity and thus the escape fraction, and the effect is stronger at lower redshifts. 

Between redshift $9$ and $6$ the scaling of the escape fraction with mass changes drastically. This is caused by the uniform UV background that in the simulation comes into full effect at redshift $9$. In haloes with virial mass $\Mvir \lesssim 10^{8} \, \Msun$ the escape fraction is very high ($\fEsc \gtrsim 0.2$) because most of the gas in these haloes is ionized and heated by the background. In this mass range there are very few haloes with zero escape fraction. There are no young stellar populations present because the star formation rate is too low in the heated gas. The UV background also affects the haloes more massive than $10^{8} \, \Msun$, although the escape fraction still drops below $10 \%$ in this mass range. Similarly, we see an escape fraction above $10 \%$ in haloes with low stellar mass ($\Mstar \lesssim 10^{4} \, \Msun$) and low gass mass ($\Mgas \lesssim 10^{6} \, \Msun$) at redshift $6$ due to the effect of the UV background. Haloes with higher stellar mass and gas mass are affected less due to their ability to self-shield.

\subsubsection{Star formation rate}\label{sec:sfr}

\begin{figure*} 
  \includegraphics[width=\textwidth]{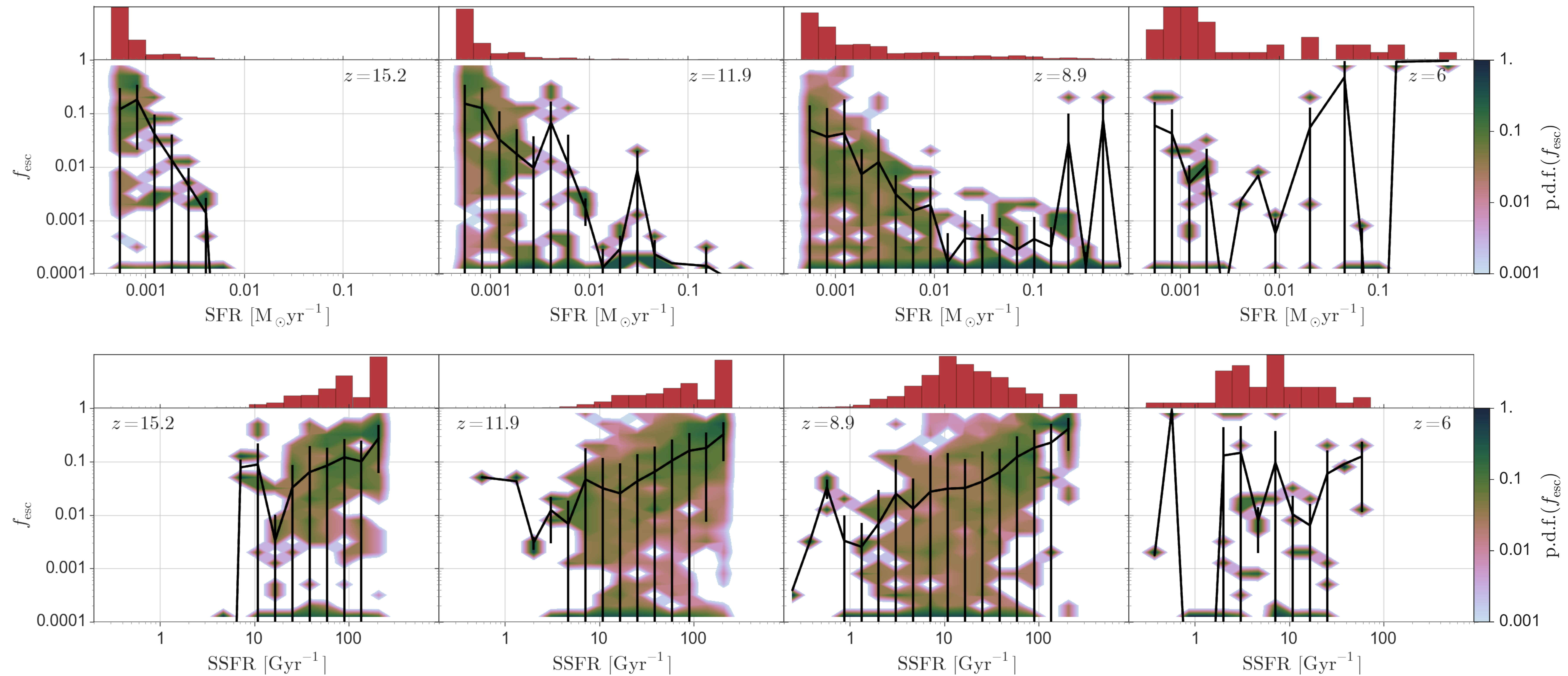}
  \caption{The escape fraction as a function of star formation rate (top) and specific star formation rate (bottom) for different redshifts. Colour-coded is the 2-dimensional histogram of all the haloes within a bin, the solid line represents the mean escape fraction in the same bin and the dashed line the mean escape fraction for haloes containing at least one stellar particle with age equal to or less than 5 Myr. The errors are the standard deviation in the mean. For clarity, haloes with $\fEsc < 10^{-4}$ are plotted at $\fEsc = 10^{-4}$. The histogram on the top shows the number of haloes in each bin. At redshift $6$ the data is from the FiBY\_S simulation only, resulting in larger Poisson errors than the higher redshifts.}
  \label{fig:fEsc_sfr}
\end{figure*}


The relation between star formation rate and escape fraction is two-fold. On the one hand, a high star formation rate means that there are many ionizing photons produced that can penetrate high column densities (see \autoref{fig:fEsc_NH} ). It also means that there is a higher rate of supernovae occurring and this feedback makes it easier for radiation to escape by redistributing the dense gas in the halo. On the other hand, a high star formation rate implies that there is cool dense gas in the halo that is likely to absorb many ionizing photons.

There are two different ways of estimating the star formation rate in the simulation. We can compute the star formation rate in every gas particle from the density and temperature, and average this over all the gas particles in the halo. This is the instantaneous star formation rate that is used in the star formation recipe in our simulations. Although this gives a good measure of the amount of dense gas in the halo, it doesn't give any information on recent episodes of star formation and therefore doesn't necessarily scale with the number of ionizing photons produced in the halo. We therefore use a different way to estimate the star formation rate, that is more consistent with how the star formation rate is determined observationally. We compute the average star formation rate in a halo by averaging the stellar mass formed over the last $5 \, \Myr$. This has the advantage that the star formation rate is proportional to the ionizing photon production. The drawback of this method is that the fixed resolution of the star particles imposes a minimum star formation rate of $2.5 \times 10^{-4} \, \Msunyr$. We have verified that the two ways of measuring the star formation rate are consistent in massive haloes in which continuous star formation takes place.

In \autoref{fig:fEsc_sfr} we show the escape fraction as a function of star formation rate and specific star formation rate. The star formation rate in the haloes ranges from $0$ (in haloes that didn't form a star particle in the last $5 \, \Myr$) to $2.3 \, \Msunyr$. We find that the average escape fraction is around $10 \%$ for all haloes with star formation rate below $10^{-3} \, \Msunyr$. This reflects the lack of dense gas in these haloes. At every redshift, around $5 \%$ of all haloes have star formation rates higher than $10^{-3} \, \Msunyr$. These haloes are massive enough to retain the star-forming gas in the presence of supernova feedback and thus have a large $\NH$. In general this results in a very low escape fraction, $\fEsc < 10^{-3}$, with only a few outliers of $\fEsc > 10^{-2}$ where feedback from a nearby stellar population has cleared away the dense gas around recently formed star particles. At redshift $6$, feedback from the UV background suppresses star formation in the majority of haloes ($86 \%$).

The decrease in escape fraction at the highest star formation rates is caused by the most massive haloes in the sample. In the previous subsection we have seen that there is a strong dependence of the escape fraction on the mass of the halo. To circumvent the mass dependence, we show the escape fraction as a function of specific star formation rate (the star formation rate divided by the stellar mass) in \autoref{fig:fEsc_sfr} as well. The specific star formation scales with the number of ionizing photons produced per solar mass in stars. It ranges from zero to $200 \, \pGyr$ in our halo sample. Haloes with a specific star formation rate higher than $50 \, \pGyr$ have escape fractions above $10 \%$. These are haloes that have recently formed most of their stars and therefore have the highest ionizing photon production per solar mass in stars. Due to the high number of ionizing photons, the young stars can ionize the dense gas around the sources and escape more easily. Haloes with specific star formation rate below $50 \, \pGyr$ show average escape fractions below $10 \%$, because the number of ionizing photons produced per solar mass is not sufficient to ionize the immediate surroundings of the young stars.

\subsubsection{Age, baryon fraction and star formation efficiency}

\begin{figure*} 
  \includegraphics[width=\textwidth]{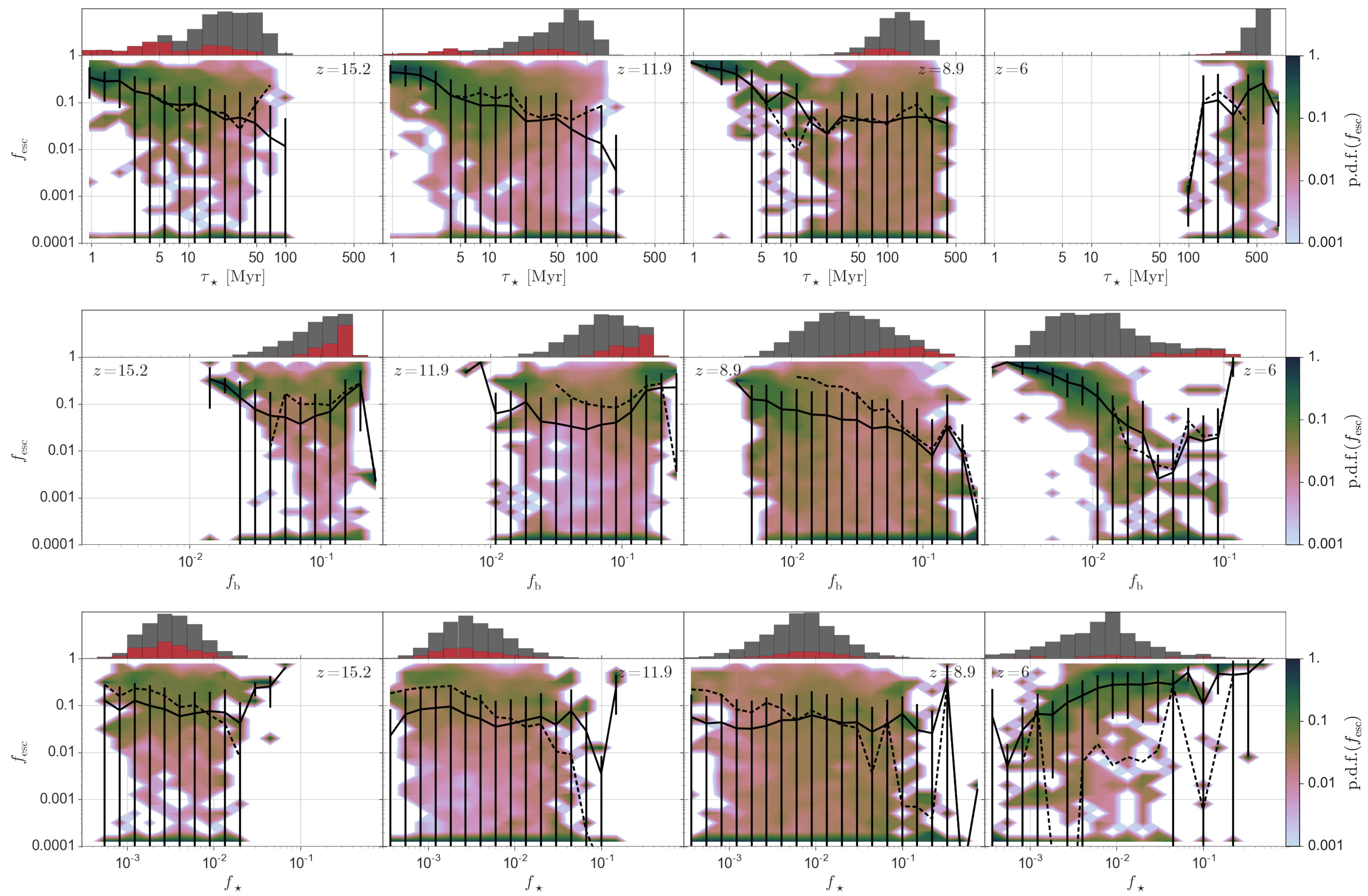}
  \caption{The escape fraction as a function of average stellar age (top), baryon fraction ($f_{\textnormal{b}} = \frac{\Mstar + \Mgas}{\Mvir}$) (middle), and star formation efficiency ($\fStar = \frac{\Mstar}{\Mgas}$) for different redshifts. Colour-coded is the 2-dimensional histogram of all the haloes within a bin, the solid line represents the mean escape fraction in the same bin and the dashed line the mean escape fraction for haloes containing at least one stellar particle with age equal to or less than 5 Myr. The errors are the standard deviation in the mean. For clarity, haloes with $\fEsc < 10^{-4}$ are plotted at $\fEsc = 10^{-4}$. The histogram on the top shows the number of haloes in each bin (grey) and the number of haloes containing at least one star particle younger than $5 \, \Myr$ (red). At redshift $6$ the data is from the FiBY\_S simulation only, resulting in larger Poisson errors than the higher redshifts.}
  \label{fig:fEsc_other}
\end{figure*}

The dense gas that is responsible for the absorption of ionizing photons in the halo is mainly redistributed by supernova feedback. Therefore the onset of the first feedback events, and thus the age of the halo, could be an important factor for the escape fraction. On the other hand, young stars emit orders of magnitude more ionizing photons, making it easier to penetrate high gas column densities which is especially important in low-mass haloes. Here we define the age of the halo as the average age of the stellar population in the halo, which ranges from $0.01 \, \Myr$ to $750 \, \Myr$.

In the top panel of \autoref{fig:fEsc_other} we show the escape fraction as a function of the average age of the star particles in the halo. At $z>6$, the escape fraction is the highest, $20 - 30 \%$, for haloes with an average age below $5 \, \Myr$. These are low-mass haloes ($\sim 10^{7} \, \Msun$) that in the last $5 \, \Myr$ underwent the first episode of star formation, hence the young average age. In most cases the gas column densities in these low-mass haloes are not high enough to shield the large number of ionizing photons produced by the young stellar populations. At these young ages there are therefore only a few haloes with zero escape fractions. Going to higher average age, the escape fraction drops to below $10 \%$ for haloes with an age larger than $5 \, \Myr$. We do not find an upward trend of the escape fraction around an age of $30 \, \Myr$ when the Pop II supernova feedback commences. In our sample the decrease in ionizing emissivity of the stellar population after $5 \, \Myr$ is more important than the onset of supernova feedback. The dashed line shows that there is no clear difference between haloes with and without young stars, although at redshift $11.9$ the latter sample shows on average a slightly higher escape fraction. This is not the case at the other redshifts. At redshift 6 there are no haloes with average age below $100 \, \Myr$ due to the effects of the UV background. Star formation in low-mass haloes is suppressed by the ionizing background. The only haloes that are still able to form stars in a reionized Universe are massive haloes that have formed stars continuously in the past and therefore have high average ages. 

The importance of the amount of dense gas in the halo for shielding of the ionizing radiation makes the baryon fraction, defined as $f_{\textnormal{b}} = \frac{\Mstar + \Mgas}{\Mvir}$, a potentially interesting quantity to constrain the escape fraction. For example, supernova feedback would lower the baryon fraction, making it easier for radiation to escape. In our halo sample the baryon fraction ranges from $0.001$ to $0.3$. However, there are very few haloes with a baryon fraction above the cosmic mean $\frac{\Omega_{\textnormal{b}}}{\Omega_{\textnormal{M}} } = 0.17$.

In the middle panel of \autoref{fig:fEsc_other} we show the escape fraction as a function of the baryon fraction. At every redshift, haloes with the lowest baryon fractions show high ($> 10 \%$) escape fractions, as expected. However, the exact baryon fraction that results in a high escape fraction changes with redshift. At $z=15.2$ a baryon fraction of $0.015$ results in an average escape fraction of $\sim 0.3$. On the other hand, at $z=8.9$ the same baryon fraction yields an average escape fraction of $\sim 0.05$. In general, at all redshifts a relatively low baryon fraction means that the star formation rate is close to zero and thus the escape fraction is high: the dense gas has been cleared by feedback and no longer shields the ionizing radiation. Because the star formation rate is so low, the ionizing photon production in these haloes is very small. 

At redshifts $15.2$ and $11.9$ there is a population of haloes with high baryon fraction and high escape fraction. These are low-mass haloes ($\lesssim 10^{7} \, \Msun$) that in the last $5 \, \Myr$ underwent the first episode of star formation. Supernova feedback has not yet cleared away any gas in the halo, hence the high baryon fraction. However, due to their low mass, the gas column density is low and therefore the escape fraction is high.

Recently, \citet{2014MNRAS.442.2560W} reported a strong trend of the escape fraction with the star formation efficiency (defined as $\fStar = \frac{\Mstar}{\Mgas}$), with the escape fraction rising from two percent at the lowest star formation efficiencies ($\fStar \sim 10^{-4}$) to nearly unity at the highest star formation efficiencies ($\fStar \sim 10^{-1}$). The reason for this is that a high $\fStar$ implies a high ionizing-photon-to-baryon ratio, thus making it easier for the photons to ionize the inter-stellar medium and subsequently escape.

In the FiBY simulations there is a strong correlation between stellar mass and gas mass in the haloes (Khochfar et al in prep.), so $\fStar$ is a measure of the scatter around this relation. For most haloes, the cause of the scatter are processes like accretion and feedback that cause the gas mass to differ from the average for a given stellar mass. However, at the low-mass end the stellar mass is limited by the SPH particle mass. Therefore, due to the probabilistic implementation of star formation, in haloes with the lowest gas masses that form only a single star particle, the stellar mass can be above the average gas mass - stellar mass relation, but never below. In these haloes we will therefore overestimate $\fStar$. In the simulations $\fStar$ ranges from $10^{-4}$ to $0.6$. We find the highest $\fStar$ at the lowest redshifts, because it takes time to grow the stellar mass needed for such a high $\fStar$.

In the bottom panel of \autoref{fig:fEsc_other} we show the escape fraction as a function of $\fStar$. We don't reproduce the rising trend of the escape fraction reported by \citet{2014MNRAS.442.2560W}. Instead, the average escape fraction is constant over the range of $\fStar$ for $15 \gtrsim z \gtrsim 9$. Exceptions are a few haloes at redshifts $15.2$ and $11.9$ with very high $\fStar$ and high $\fEsc$. In these cases the high $\fStar$ is caused by the overestimate of $\fStar$ in haloes with low gas mass, which in turn also leads to a high escape fraction. However, this happens only in 4 haloes at each of these two redshift bins. At redshift $6$ the trend is different due to the effect of the UV-background. The UV-background has the largest effect in haloes with low gas mass, that is, with high $\fStar$ at given stellar mass. In these haloes the escape fraction is therefore higher compared to higher redshifts.

The reason we find no trend of the escape fraction with $\fStar$ at $15 \gtrsim z \gtrsim 9$ is that $\fStar$ is not an accurate probe of the number of ionizing photons per baryon in the halo. For example, depending on the age of the stellar population, a high $\fStar$ in a low-mass halo can either mean a high number of ionizing photons per baryon in a halo that has recently formed a star or a low number of ionizing photons in a halo hosting an old stellar population that blew away most of the gas when the massive stars ended their lives as supernovae. We find that the specific star formation rate is a more accurate probe of the number of ionizing photons produced if the star formation rate is averaged over the last $5 \, \Myr$ (see \autoref{sec:sfr}).

In a previous study of high-redshift dwarf galaxies we found a strong correlation between the escape fraction and the spin of the halo \citep{Paardekooper:2011cz}. We do not reproduce this result for the proto-galaxies in the FiBY simulations and find no correlation between the escape fraction and spin of the halo at all. The main reason for this is the mass and morphology of the isolated galaxies in our previous study: they were more massive and we assumed that they had formed a rotationally supported disc. In that case the spin of the halo is important for the effectiveness of supernova feedback. On the other hand, most of the proto-galaxies we follow in the FiBY have a much more irregular morphology, which makes the spin of the halo of much less importance.

\subsection{Principle component analysis of the halo properties}\label{sec:pca}

In the previous section we have studied the dependence of the escape fraction on several properties of the host halo. In all cases there was a significant scatter around the average relation with the escape fraction. Using only one of these properties for predicting the escape fraction in a halo will therefore lead to inaccurate results as this scatter is neglected. Ideally we would like to describe the escape fraction as a function of only 
one or two variables. Because we have a large data sample, 75683 haloes in total, we can use principal component analysis (PCA) to study whether we can reduce the number of variables in the data. The principal components (PCs) are computed by diagonalising the covariance matrix of the data\footnote{In practice a more stable way to compute the principal components is to compute the single value decomposition of the data matrix.}. The eigenvectors of the covariance matrix are the principle components, while the eigenvalues corresponding to the eigenvectors represent the relative importance of the principal components.


We have performed PCA on the data discussed in \autoref{sec:halo_properties}: redshift $z$, virial mass $\Mvir$, stellar mass $\Mstar$, gas mass $\Mgas$, star formation rate, specific star formation rate, average stellar age $\tau_{\star}$, baryon fraction $f_{\textnormal{b}}$, and star formation efficiency $\fStar$. To account for the large dynamic range we have used the logarithm of all parameters, and all variables were standardised (that is, transformed to have mean zero and variance one) before performing the PCA. Three of the parameters, the specific star formation rate, baryon fraction and star formation efficiency are derived from the other parameters, something the PCA should reflect.

\begin{figure} 
  \includegraphics[width=85mm]{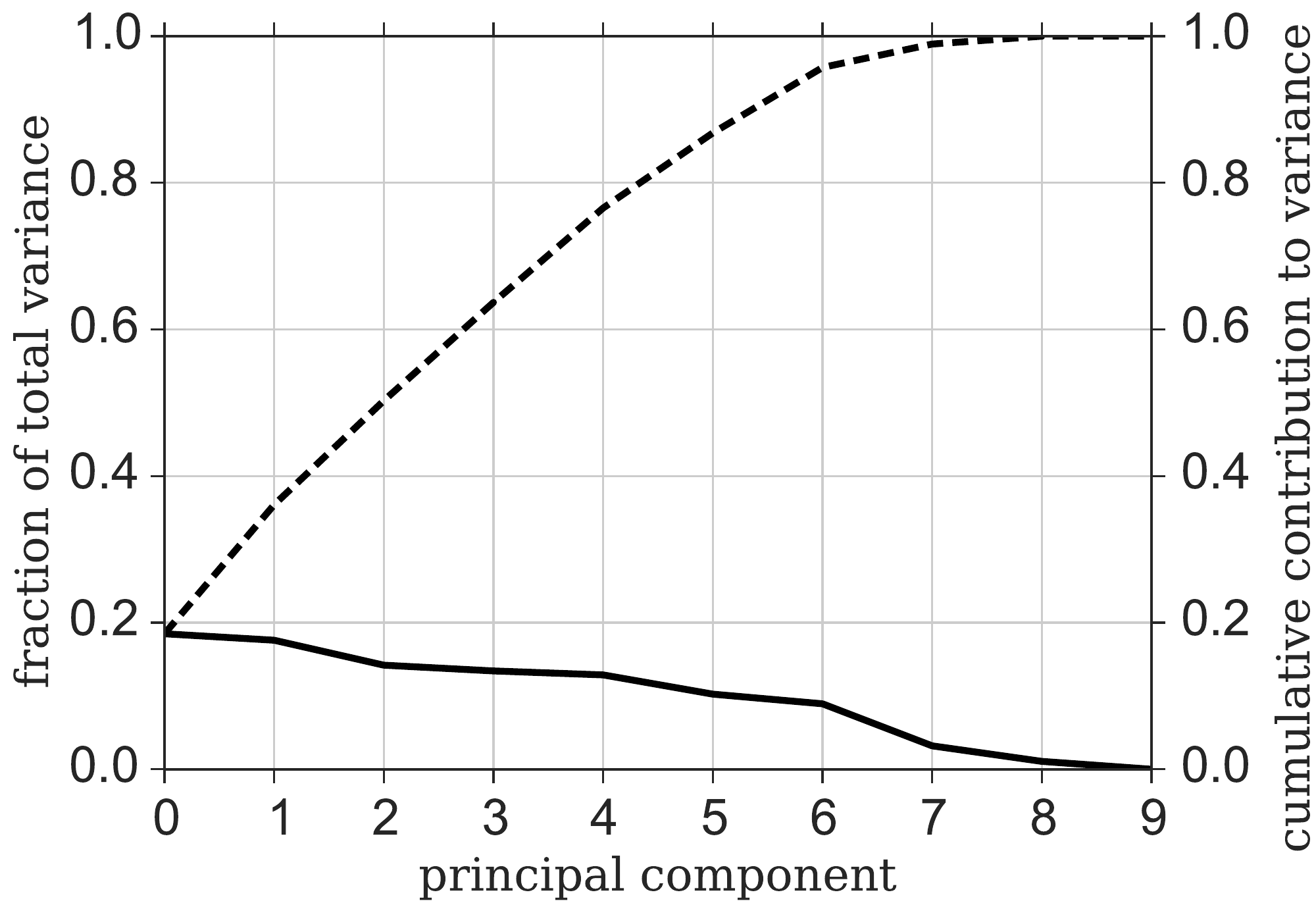}
  \caption{ 
Fraction of the variance in each principal component (solid line) and cumulative contribution of the principal components (dotted line).
  }
  \label{fig:pca_eigenvalues}
\end{figure}

In \autoref{fig:pca_eigenvalues} we show the contribution of the eigenvalues to the total variance of the data. The first component accounts for 18 \% of the variance, as does the second. To account for $50\%$ of the variance we need to include at least 3 PCs, whereas we need 7 PCs to account for $90\%$ of the variance. This shows that we cannot reduce the number of variables for predicting the escape fraction without losing information. As expected, the last 3 PCs account for only $4\%$ of the variance, because 3 of our variables are derived from the other variables.

\begin{figure} 
  \includegraphics[width=85mm]{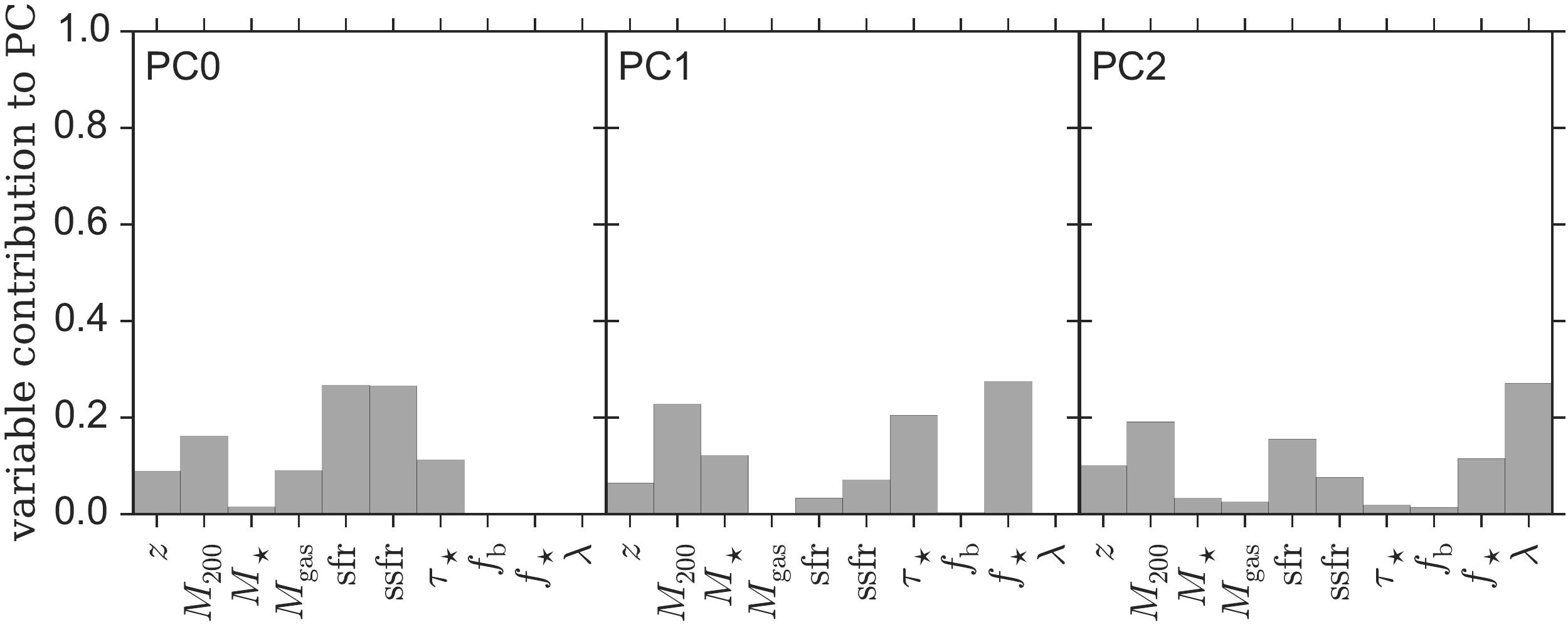}
  \caption{ 
The contribution of all considered variables to the first three principal components that account for $50\%$ of the variance.
  }
  \label{fig:pca_var_contribution}
\end{figure}

In \autoref{fig:pca_var_contribution} we show the contribution of all variables to the first three principal components: PC0, PC1 and PC2. These principal components account for $50\%$ of the variance. PC0 is dominated by the star formation rate and specific star formation rate, whereas PC1 is dominated by the mass (virial mass and $\fStar$, which is a combination of the gas mass and the stellar mass), although the average stellar age has also a strong contribution. PC3 is mostly dominated by the spin parameter $\lambda$, with contributions from the star formation rate, mass and $\fStar$ as well.

To check whether there are any strong correlations between haloes that have very high or very low escape fractions, we have performed PCA on these subsets of the data as well. This confirms that the main constraint for the escape fraction, namely the high gas column densities between the young sources and the virial radius, does not show a tight correlation with any property of the halo. We have also verified that using a more robust measure of the variance like the median absolute deviation \citep[see][]{2014MNRAS.440..240D} does not change our results much. In this case PC0 is slightly more important ($23\%$ of the variance) compared to PC1 ($15\%$ of the variance). The contribution of all other PCs is similar. In this case, $\fStar$ does not contribute much to the first 3 PCs, instead the baryon fraction becomes more important. All other variables have similar contributions to the first 3 PCs.

\subsection{The probability distribution of the escape fraction}

The results in the previous sections show that is very hard to predict the escape fraction of an individual halo. Even though there are general trends like higher escape fractions in lower mass haloes, and higher escape fractions in haloes with higher specific star formation rates, the scatter around these relations is so large that haloes with similar properties can exhibit a large range of escape fractions. The origin of this scatter is physical in nature and has to do with the distribution of dense gas in the halo that absorbs most of the ionizing photons. For this reason it is useful to look at the escape fraction in a different way: given a halo with certain properties, what is the probability that this halo has a certain escape fraction?

\begin{figure*} 
  \includegraphics[width=\textwidth]{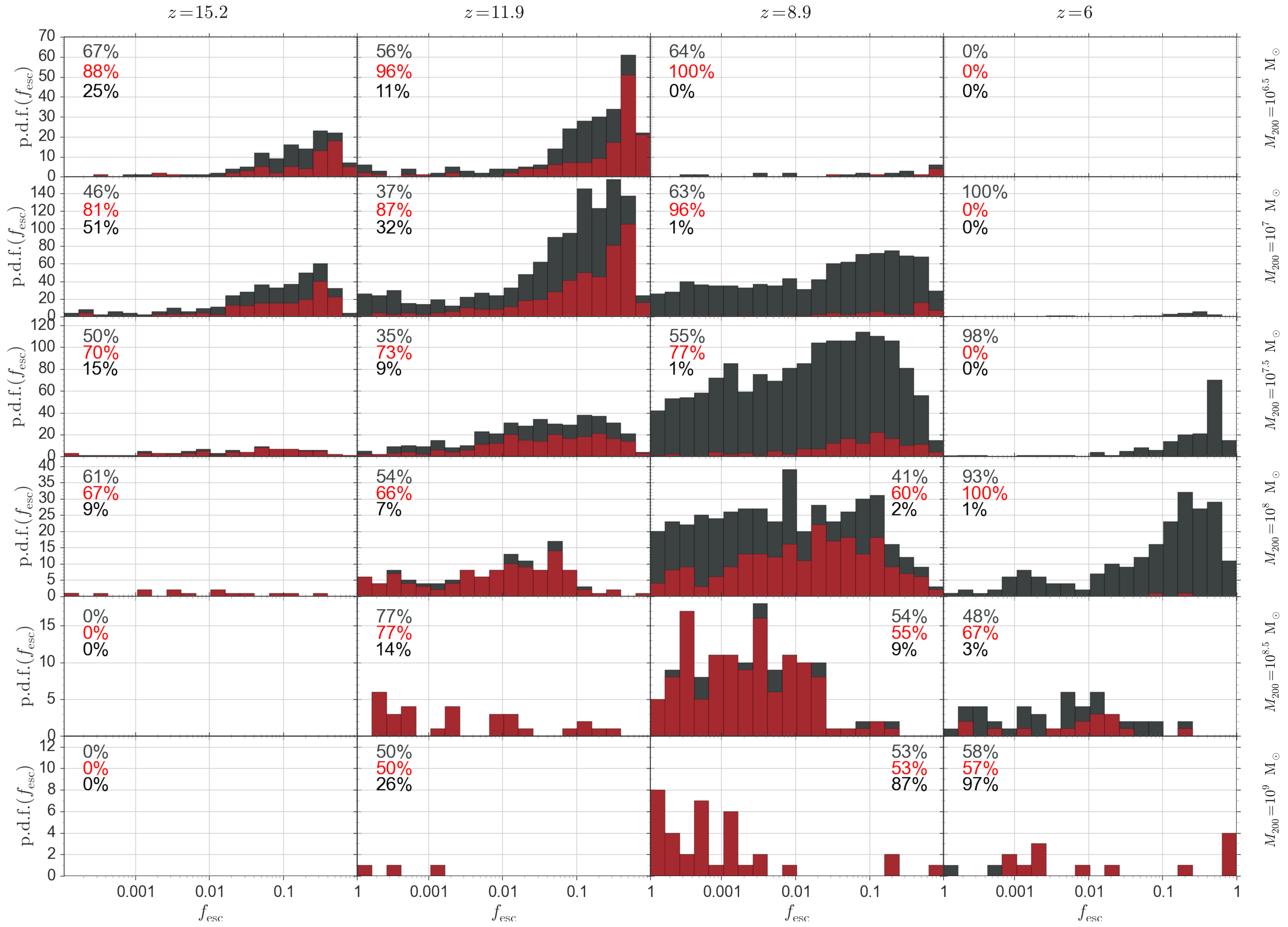}
  \caption{ The probability density function of the escape fraction in 6 different mass bins, at 4 different redshifts, for all haloes (grey) and haloes containing at least one star particle younger than $5 \, \Myr$ (red). The bins are evenly spaced in log, the width of each bin is 0.5 dex. The first two percentages represent the fraction of haloes with $\fEsc \ge 10^{-4}$, i.e. the percentages of haloes shown in the plots, for all haloes (top) and haloes containing at least one star particle younger than $5 \, \Myr$ (middle). The third percentage is the fraction of the total ionizing photons that is produced by the stars in the halo mass bin. At redshift $6$ the data is from the FiBY\_S simulation only, resulting in fewer haloes than the higher redshifts.}
  \label{fig:pdf_fEsc_mVir}
\end{figure*}

In \autoref{fig:pdf_fEsc_mVir} we show as an example the probability density of the escape fraction for haloes subdivided in mass bins of 1 dex. The top row shows haloes in the lowest mass range that form stars in our simulations: mini-haloes with mass around $5 \times 10^6 \, \Msun$. At redshifts higher than 12, in the mini-haloes, the distribution of the escape fraction is peaked around $\fEsc \approx 0.5$. The gas mass of these haloes is not very high, hence it is relatively easy for radiation to escape. Haloes that host a stellar population younger than $5 \, \Myr$ tend to have higher escape fraction, because the high ionizing photon production of these stars means that most of the gas in the halo is ionized. Moreover, more than $80 \%$ of these star-forming mini-haloes have escape fractions higher than $0.01$, showing that these haloes are efficient contributors to reionization. In general these mini-haloes from only a single generation of stars before star formation is halted by internal feedback so their contribution is very short. However, these haloes are abundant enough that stars formed in mini-haloes contribute $25 \%$ of the ionizing photon budget at $z=15$, and $11 \%$ at $z=12$. Due to the high escape fractions, many of the produced photons are available for reionization. After the UV-background turns on in the simulation, these haloes rapidly become unimportant because their gas is photo-evaporated.

Haloes with ten times higher masses, $5 \times 10^7 \, \Msun$ (third row in \autoref{fig:pdf_fEsc_mVir}), are able to host multiple stellar populations and therefore contribute continuously to reionization. Escape fractions in these haloes are generally lower than in mini-haloes. The distribution at $z=15$ is flat, but evolves towards higher escape fractions at lower redshifts, with peaks around $\fEsc \approx 0.05$ at $z=12$ and $\fEsc \approx 0.15$ at $z=9$. Similar to the mini-haloes, haloes supporting stellar populations younger than $5 \, \Myr$ show a preference for larger escape fractions. However, only about $50 \%$ of the star-forming haloes in this mass range have escape fractions higher than $0.01$, the larger gas content compared to mini-haloes makes it harder for the ionizing photons to escape. At redshift 6 haloes in this mass range can no longer form stars due to the effect of the UV-background. Before the UV-background comes into effect, stars in haloes with $\Mvir \le 5 \times 10^{7} \, \Msun$ contribute the majority of ionizing photons to the total budget. Combined with the high escape fractions this stresses the importance of haloes in this mass range for reionization.

We find the same trend in haloes with masses of $1 \times 10^8 \, \Msun$ (fourth row in \autoref{fig:pdf_fEsc_mVir}): the distribution starts out flat and evolves towards a peaked distribution. Compared to the lower mass range, the shift towards a peaked distribution happens at lower redshifts ($z=9$), and the peak is at lower values of the escape fraction ($\fEsc \approx 0.02$). Only around $25 \%$ of the haloes hosting young stellar populations have $\fEsc > 0.01$. Although individual haloes in this mass range form more stars than their lower-mass counterparts, they are not abundant enough to produce more than $10 \%$ of the total number of ionizing photons produced. Combined with the lower escape fractions this shows that these haloes are less efficient sources of reionization compared to lower-mass haloes. At $z=6$ most of the star formation is quenched by the UV-background.

The highest mass bins we consider here are $5 \times 10^8 \, \Msun$ and $1 \times 10^9 \, \Msun$ (fifth and sixth row in \autoref{fig:pdf_fEsc_mVir}, respectively). It is very hard for ionizing photons to escape from these haloes due to the high column densities around the sources (see \autoref{fig:NH_mVir}). Very few haloes in this mass range have escape fractions higher than $10 \%$ and the distribution is peaked towards escape fractions much lower than that. Moreover, at $z=9$ less than $10\%$ of the haloes in this mass range have $\fEsc > 0.01$. This shows why the most massive haloes in our sample are inefficient contributors to reionization: even though the stars in these haloes produce many ionizing photons (at $z=12$ about $40 \%$ of the ionizing photons are produced in haloes with $\Mvir \ge 2 \times 10^{8} \, \Msun$, rising to more than $90 \%$ at $z=9$ when the UV-background quenches the lower-mass haloes), the vast majority of the produced ionizing photons are absorbed in the haloes themselves.

\begin{figure} 
  \includegraphics[width=85mm]{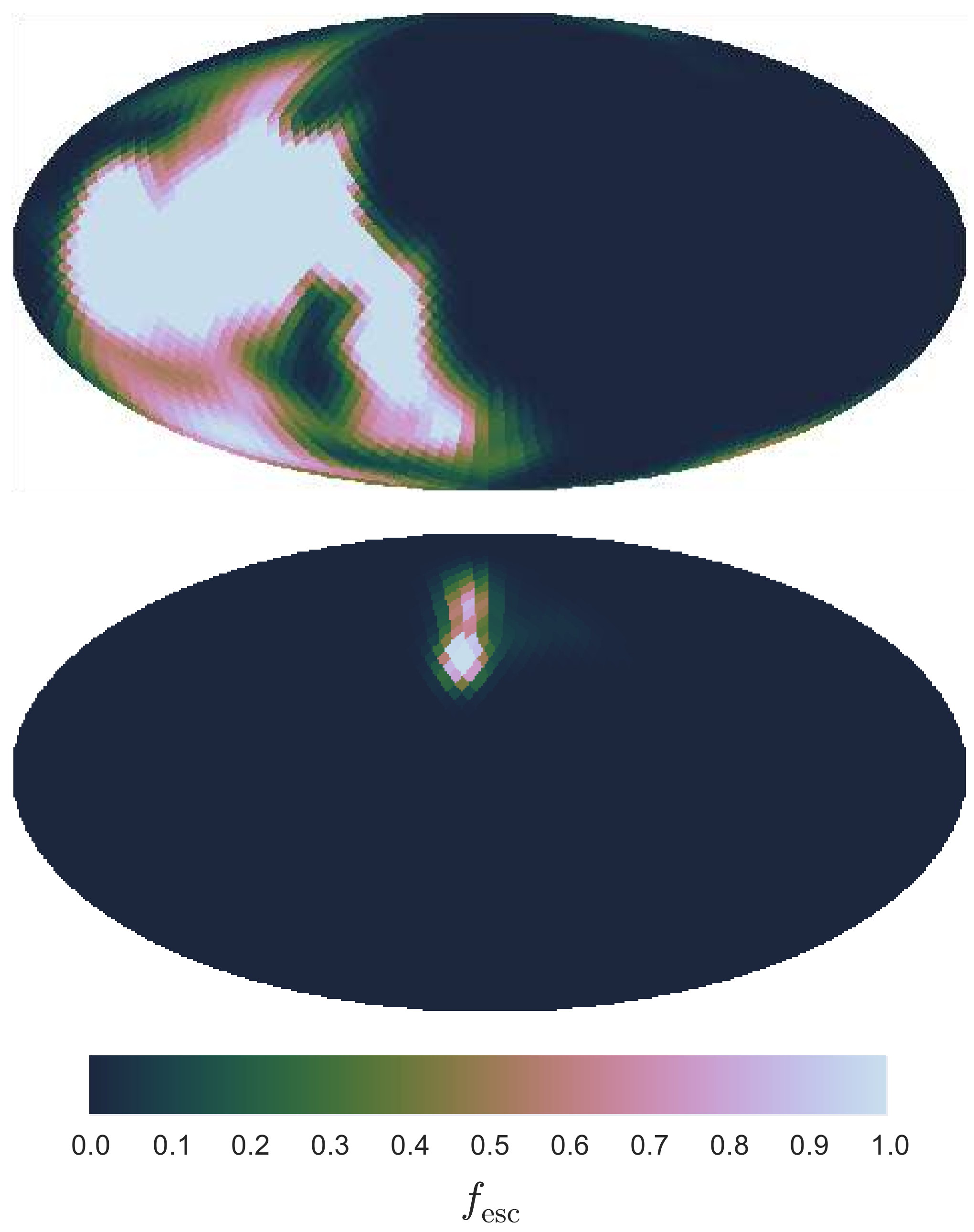}
  \caption{The angular distribution of the escape fraction in a sphere at the virial radius of two haloes at $z=12$. Both haloes have $\Mvir \approx 6 \times 10^{7} \, \Msun$, specific star formation rate $\approx 40 \, \pGyr$, $f_{\textnormal{b}} \approx 0.08$ and $\tau_{\star} \approx 44 \, \Myr$. The halo in the upper panel has an average escape fraction of $0.3$, while the halo in the bottom panel has an average escape fraction of $0.01$.
  }
  \label{fig:all_sky_similar}
\end{figure}

Our analysis shows that assuming a single value for the escape fraction, as is done in many studies of reionization, or assuming for example a mass-dependent escape fraction, does not do justice the large scatter that we find in our simulations. The escape fraction is much better represented by the probability distributions we show in \autoref{fig:pdf_fEsc_mVir}. The origin of this scatter does not lie in other halo properties like specific star formation rate, baryon fraction or age, but rather in the distribution of dense gas in the halo. To emphasise this point we show in \autoref{fig:all_sky_similar} two haloes with similar properties: $\Mvir \approx 6 \times 10^{7} \, \Msun$, specific star formation rate $\approx 40 \, \pGyr$, $f_{\textnormal{b}} \approx 0.08$ and $\tau_{\star} \approx 44 \, \Myr$. Despite these similarities, the escape fractions differ by more than one order of magnitude: the halo in the upper panel has an average escape fraction of $0.3$, while the halo in the bottom panel has an average escape fraction of $0.01$. This is only due to the distribution of dense gas. In the halo with high escape fraction there is a large hole in the dense gas through which ionizing photons can escape, while in the halo with low escape fraction there is only a very small channel in which the ionizing photons are not absorbed.

\subsection{The angular dependence of the escape fraction}

In \autoref{sec:absorption} we have shown that the amount and distribution of dense gas is the main constraint on the escape of ionizing photons. A natural consequence is that ionizing radiation escapes primarily through channels with low column density, hence the escape fraction is highly anisotropic. How anisotropic the escape fraction is, is very important for observational studies of the escape fraction, because it tells us the chance that a galaxy emits ionizing radiation in our direction. It is also very important for numerical simulations of cosmic reionization, because highly anisotropic escape fractions could potentially impact the topology of reionization. 

\begin{figure*} 
  \includegraphics[width=\textwidth]{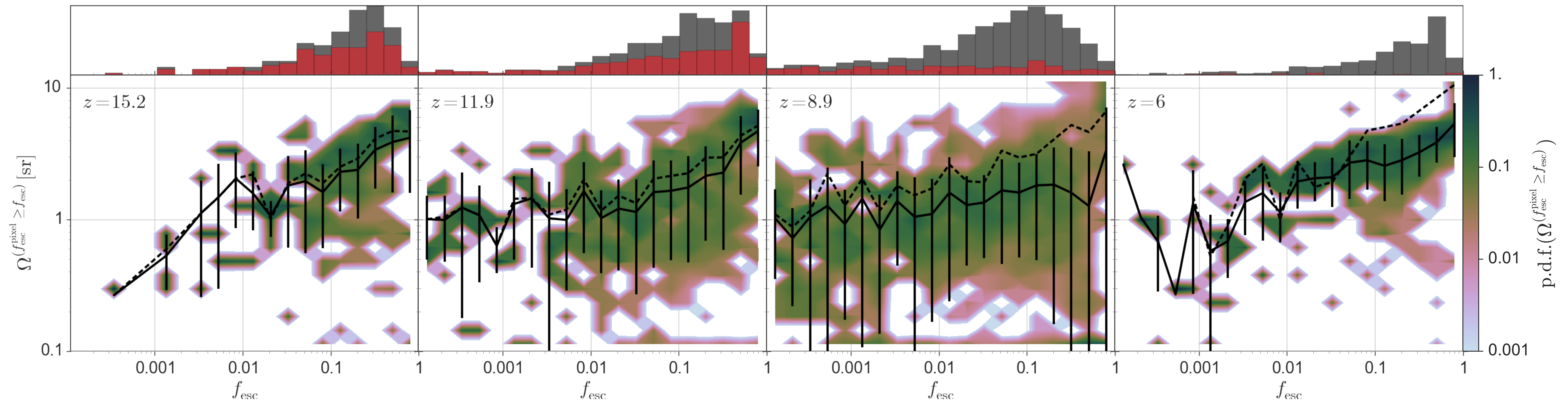}
  \caption{The solid angle where the escape fraction is equal to or larger than the average over all angles as a function of the average escape fraction for different redshifts. Colour-coded is the 2-dimensional histogram of all the haloes within a bin, the solid line represents the mean solid angle in the same bin and the dashed line the mean solid angle for haloes containing at least one stellar particle with age equal to or less than 5 Myr. The errors are the standard deviation in the mean. For clarity, haloes with $\Omega(\fEsc) < 0.1$ are plotted at $\Omega(\fEsc) = 0.1$. The histogram on the top shows the number of haloes in each bin (grey) and the number of haloes containing at least one star particle younger than $5 \, \Myr$ (red). At redshift $6$ the data is from the FiBY\_S simulation only, resulting in larger Poisson errors than the higher redshifts.}
  \label{fig:SA_fEsc}
\end{figure*}

We study the anisotropy of the escape fraction by determining in every halo the solid angle $\SA$ on the sky in which the escape fraction is equal to or higher than the escape fraction averaged over the entire sky. We do this by computing the fraction of pixels in the HealPix map (computed as explained in \autoref{sec:ray_trace}) in which the average escape fraction is higher than or equal to the escape fraction averaged over all pixels. This is done for individual pixels independently of all others, the total solid angle can therefore consist of many individual patches in which the escape fraction is higher than average. To avoid numerical artefacts due to the different computations of the escape fractions (see \autoref{sec:ray_trace}) we exclude all haloes in which the average escape fraction in the all-sky maps is less than $10^{-10}$.

In \autoref{fig:SA_fEsc} we show $\SA$ as a function of $\fEsc$. On average, a higher escape fraction means a larger solid angle on the sky in which radiation escapes. However, there is a large scatter that has to do with how the high-column density gas is distributed. In haloes with similar escape fraction there can be a few sightlines that have very low column density through which the radiation escapes or there can be many sightlines of slightly higher column density each. In haloes with $\fEsc > 0.5$, on average $\SA \sim 2\pi \, \sr$. For lower escape fractions the solid angle is smaller, dropping to $\SA \sim 0.3 \pi \, \sr$ for $\fEsc = 0.01$.  

We find that there is a clear distinction between haloes that host at least one stellar population younger than $5 \, \Myr$ and haloes that contain only old stars. At fixed escape fraction, $\SA$ is larger in the haloes hosting a young stellar population. because the young massive stars are so luminous that they penetrate higher column densities. That means that on average there are more sight lines through which radiation can escape. At redshift $9$ and higher, the apparent evolution of the average $\SA$ is caused by the growth of the population of haloes without young stellar populations with respect to haloes that actively form stars. For haloes that host at least one young stellar population the average relation between $\SA$ and $\fEsc$ does not change with redshift (dashed line in \autoref{fig:SA_fEsc}). However, at redshift $9$ the population of haloes that have not formed stars in the last $5 \, \Myr$ has grown relative to the population with young stars and at fixed $\fEsc$, $\SA$ is lower in these haloes. This causes the apparent evolution of the average $\SA$ from $z=15$ to $z=9$. This changes at $z=6$, when the UV-background heats up the gas in low-mass haloes. In low-mass haloes the escape of radiation becomes much more uniform because they are photo-evaporated by the UV-background. However, haloes that are still actively forming stars show on average the same relation as haloes at higher redshifts.


\begin{figure*} 
  \includegraphics[width=\textwidth]{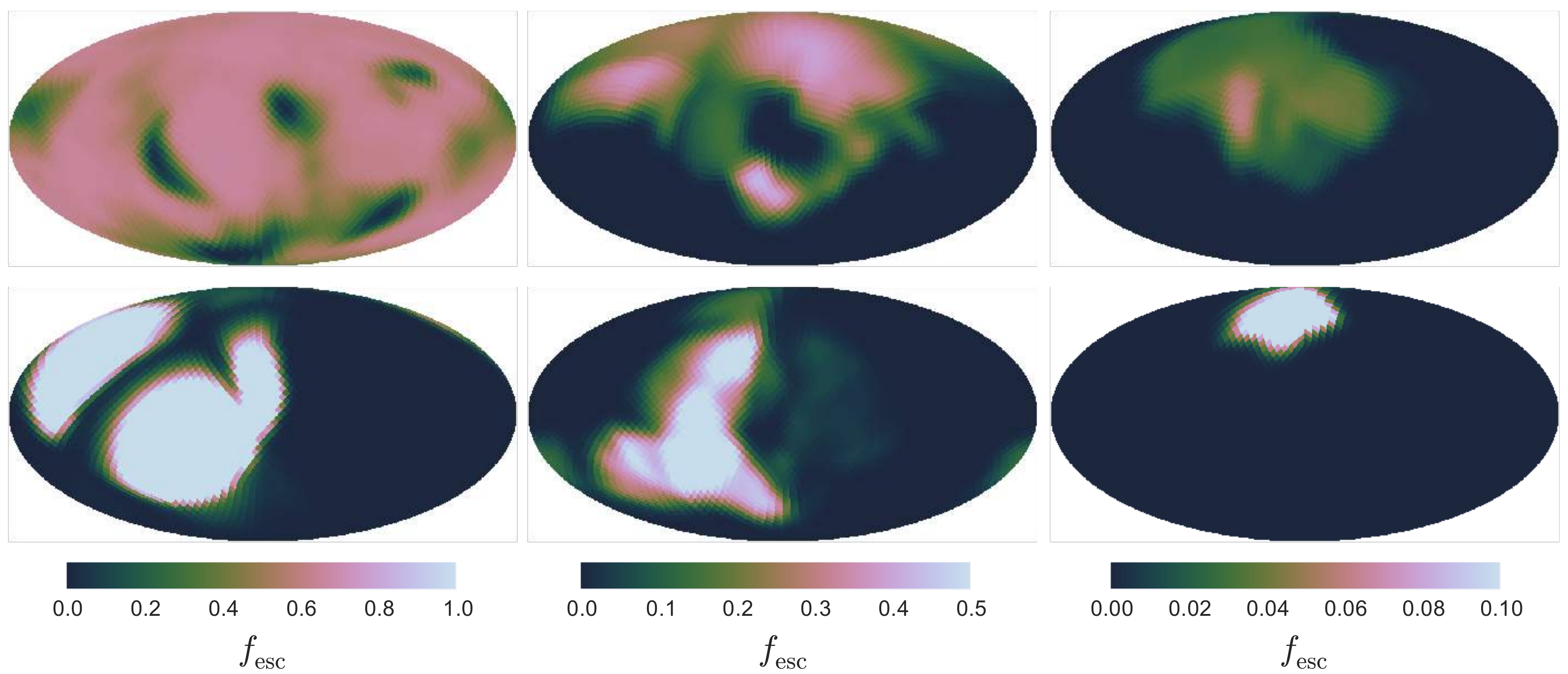}
  \caption{The angular distribution of the escape fraction in a sphere at the virial radius of six haloes at $z=12$. The left column shows two haloes with $\fEsc \approx 0.6$, the halo in the top panel has $\SA = 2.8\pi \, \sr$ and the halo in the bottom panel has $\SA = \pi \, \sr$. The middle  column shows two haloes with $\fEsc \approx 0.1$, the halo in the top panel has $\SA = 1.7\pi \, \sr$ and the halo in the bottom panel has $\SA = \pi \, \sr$. The right column shows two haloes with $\fEsc \approx 0.01$, the halo in the top panel has $\SA = 1.1\pi \, \sr$ and the halo in the bottom panel has $\SA = 0.2\pi \, \sr$.}
  \label{fig:all_sky}
\end{figure*}

\autoref{fig:SA_fEsc} shows that there is a relation between the escape fraction and the solid angle on the sky in which the escape fraction is equal to or higher than the angle-averaged escape fraction. This figure doesn't tell us whether the radiation escapes through several small channel or that there is one large solid angle on the sky through which most radiation escapes. In \autoref{fig:all_sky} we show all-sky maps of the escape fraction for six haloes that are representative for haloes with high ($\fEsc \approx 0.5$; left column), medium ($\fEsc \approx 0.1$; middle column) and low ($\fEsc \approx 0.01$; right column), which show escape through either large (top row) or small (bottom row) solid angle. In almost all cases that we studied the ionizing radiation escapes through one or two patches on the sky in which the escape fraction is much higher than the angle average, in all other directions the escape fraction is close to zero. At fixed escape fraction a higher $\SA$ in general means that the patch through which the radiation escapes is larger and that the escape fraction in the patch is lower (the difference between the top and bottom row in \autoref{fig:all_sky}), not that there are more distinct patches.

\subsection{The spectrum of outgoing radiation}

The frequency dependence of the escape fraction is something that is not well studied. Most recent simulations transport photons in only a single frequency bin for computational efficiency. While this is sufficient to compute the escape fraction of hydrogen-ionizing photons when only absorption by hydrogen atoms is taken into account, it neglects the influence of helium. In our simulations we employ 10 frequency bins and take into account absorptions by both hydrogen and helium to study the frequency dependence of the escape fraction. This is important for example for studying the start of helium reionization, because the most massive stars (especially Pop III stars) emit photons energetic enough to ionize $\HeII$. This in turn is important for the temperature state of the inter-galactic medium \citep{2012MNRAS.423..558C}. Another example is the modelling of carbon absorption systems during reionization, in which a frequency-dependent escape fraction could play an important role for modelling the ratio of \ion{C}{II} to \ion{C}{IV} \citep{Finlator:2014tw}.

\begin{figure*} 
  \includegraphics[width=\textwidth]{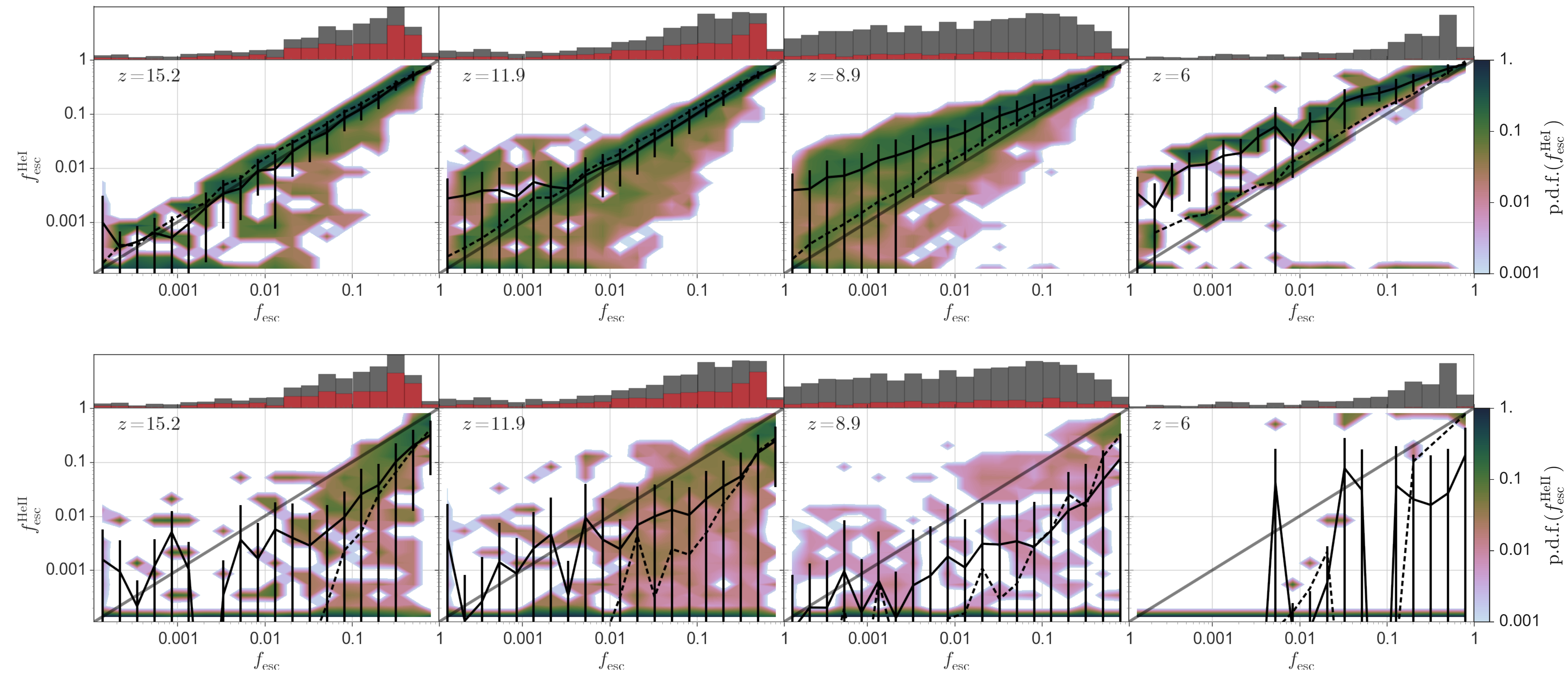}
  \caption{(top) The escape fraction of $\HeI$-ionizing photons as a function of the escape fraction of $\HI$-ionizing photons for different redshifts. (bottom) The escape fraction of $\HeII$-ionizing photons as a function of the escape fraction of $\HI$-ionizing photons for different redshifts. Colour-coded is the 2-dimensional histogram of all the haloes within a bin, the solid line represents the mean escape fraction in the same bin and the dashed line the mean escape fraction for haloes containing at least one stellar particle with age equal to or less than 5 Myr. The errors are the standard deviation in the mean. For clarity, haloes with $\fEsc < 10^{-4}$ are plotted at $\fEsc = 10^{-4}$. The histogram on the top shows the number of haloes in each bin (grey) and the number of haloes containing at least one star particle younger than $5 \, \Myr$ (red). At redshift $6$ the data is from the FiBY\_S simulation only, resulting in larger Poisson errors than the higher redshifts.}
  \label{fig:fEscHe_fEscH}
\end{figure*}

In the top panel of \autoref{fig:fEscHe_fEscH} we show the escape fraction of photons energetic enough to ionize $\HeI$ ($h \nu > 24.6 \, \eV$) as a function of the escape fraction of $\HI$-ionizing photons ($h \nu > 13.6 \, \eV$). Because the ionization energies of hydrogen and neutral helium are so close to each other, and because the spectrum of young stellar populations is flat in this range (see \autoref{fig:freqBins}), it is generally assumed that helium is singly ionized together with hydrogen. For this reason it is no surprise that the average $\fEscHeI$ is similar to the average $\fEsc$. For $\fEsc > 0.1$ this is true at all redshifts, because at such high escape fractions most hydrogen and helium in the halo is singly ionized. At redshift 15 the linear relation holds for all escape fractions, but as we go to lower redshifts two classes of haloes appear: those actively forming stars (dashed line) for which the linear relation still holds, and a population of haloes for which $\fEscHeI$ is higher than $\fEsc$. The latter haloes are of low mass ($\Mvir \approx 2 \times 10^7$) and are not actively forming stars. The old stellar populations in these haloes cannot ionize the gas, which means that $\HeI$-ionizing photons can escape more easily due to the smaller cross section of $\HeI$. However, in these haloes $\fEscHeI$ is always below $10\%$, so still not many of the produced high-energy photons make it out of the halo.

For photons of even higher energy, capable of ionizing $\HeII$ ($h \nu > 54.4 \, \eV$), the situation is different. In the bottom panel of \autoref{fig:fEscHe_fEscH} we show the escape fraction of $\HeII$-ionizing photons as a function of the escape fraction of $\HI$-ionizing photons. On average, $\fEscHeII$ is always lower than $\fEsc$. The reason is that Pop II stellar populations emit much less photons with these high energies, see \autoref{fig:freqBins}. The few photons at these energies are therefore mostly absorbed by $\HeII$ atoms. The same thing can be seen in \autoref{fig:Test2}, where the sphere in which $\HeII$ is ionized has a much smaller radius than the radii at which $\HI$ and $\HeI$ are ionized. For this reason, $\fEscHeII$ is on average smaller than $\fEsc$, independent of redshift.

\begin{figure} 
  \includegraphics[width=85mm]{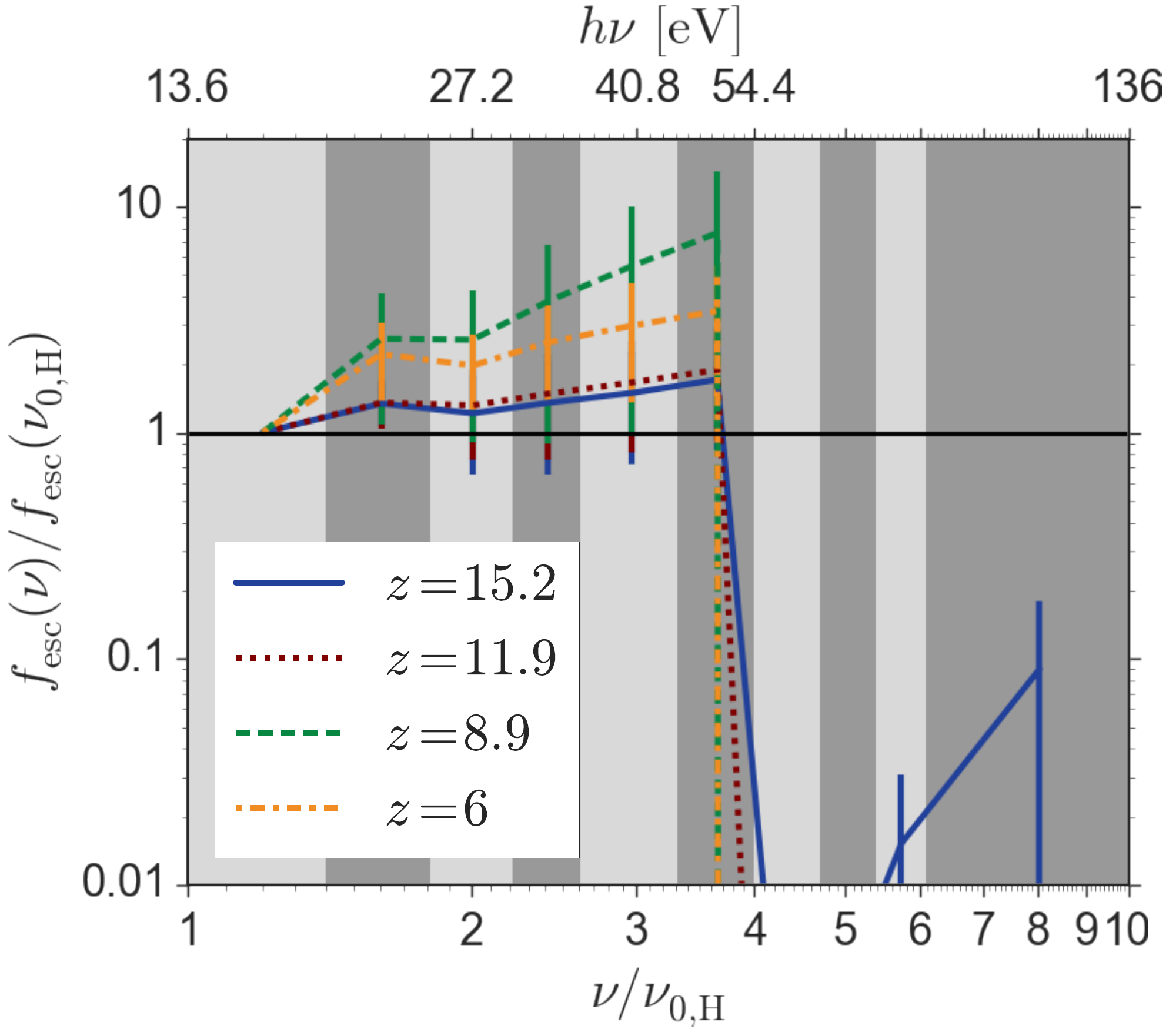}
  \caption{ 
Escape fraction of photons in every frequency bin, normalised to the escape fraction in the first bin, computed by taking the median of all haloes with escape fraction larger than zero. Error bars are the median absolute deviation. Shaded areas represent the frequency bin spacing. 
  }
  \label{fig:fEsc_spectrum}
\end{figure}

To get a more accurate picture of the frequency dependence of the escape fraction we show in \autoref{fig:fEsc_spectrum} the escape fraction in every frequency bin relative to the escape fraction in the first bin. We have only included haloes in which the escape fraction in the first bin is larger than $10^{-5}$. This figure shows the trend described previously more clearly: as we go to bins of higher frequency, the escape fraction is higher until we reach the ionization frequency of $\HeII$. If we assume that radiation escapes through sight lines that have ionized both $\HI$ and $\HeI$, this can be understood as follows. The number of absorptions at a certain frequency depends on the optical depth at that frequency. In gas with primordial composition there are roughly 12 times more hydrogen atoms than helium atoms. That means that the ratio of the optical depth at the ionization frequency of $\HeI$ to the optical depth at the ionization frequency of $\HI$ is:
\begin{equation}
    \frac{\tau(\nuHeI)}{\tau(\nuH)} = \frac{\Delta \ell \, \nH (\sigma_{\textnormal{H}}(\nuHeI) + \frac{1}{12}\sigma_{\textnormal{He I}}(\nuHeI))}{\Delta \ell \, \nH \, \sigma_{\textnormal{H}}(\nuH)} \approx 0.3,
\end{equation}     
where $\Delta \ell$ is the path length along the line of sight, $\nH$ is the number density of hydrogen and $\sigma$ is the cross section. We therefore expect that at the ionization frequency of $\HeI$, the third frequency bin, the escape fraction is about a factor of 3 higher than in the first bin. In higher frequency bins this effect is stronger and therefore the escape fraction is higher, until the ionization frequency of $\HeII$ is reached. There, the escape fraction is much lower than in the first bin, because the stellar populations do not emit many photons at these high frequencies so they are mostly absorbed by the $\HeII$. The redshift dependence of the relation is mainly caused by low-mass haloes with only old stellar populations. If we exclude those, at all redshifts the relation is similar to the one at the highest redshift.

\section{Discussion}

We have found that dense gas in the vicinity of the young massive stars is the main constraint on the escape fraction of ionizing photons from high-redshift galaxies: most ionizing photons are absorbed in a radius of $10 \, \pc$. Previous studies have shown the importance of the porosity of the inter-stellar medium for the escape fraction \citep{Ciardi:2002bz,2011ApJ...731...20F}, especially in the vicinity of the sources \citep{Paardekooper:2011cz}. Our results refine the study of \citet{Kimm:2014wh}, who found that in galaxies of low escape fraction the optical depth in a radius of $100 \, \pc$ is $\sim 2-4$, large enough to absorb the majority of produced ionizing photons. Similarly, \citet{2013ApJ...775..109K} found that in simulations of disc galaxies the average escape fraction is determined by the small-scale properties of the clouds surrounding the star particles. This has lead to the idea that feedback from supernova explosions is an important factor in the escape fraction. However, since supernova feedback sets in when the massive stars in the stellar population (that produce the majority of ionizing photons) have disappeared, the relation between supernova feedback and the escape fraction is complex.

Due to the delay between the onset of star formation and feedback, supernovae will play the largest role in haloes that are massive enough to continuously form stars. In our large sample of haloes we find that in these haloes, with $\Mvir \ge 10^8 \, \Msun$, the escape fraction is below $0.001$ unless feedback clear away the birth cloud of young star particles, or causes an inhomogeneous density distribution around the young sources. Although this confirms the importance of feedback in haloes within this mass range, the fraction of haloes with $\fEsc > 0.01$ is small, varying from redshift $15$ to $9$ between $29 \%$ and $15 \%$. This fraction gets even smaller when we consider haloes of higher masses. 

Haloes with masses smaller than $10^8 \, \Msun$ form in general one stellar population, because supernova feedback from the massive stars delays subsequent star formation. In these haloes, supernova feedback is therefore only effective after most of the ionizing photons have already been produced. Despite this we find high escape fractions in this mass range, and on average higher escape fractions in haloes in which the stellar population is younger than $5 \, \Myr$, that are unaffected by supernova feedback. The reason is that due to the smaller potential well, the gas column density in the vicinity of the sources is not high enough to shield the ionizing radiation. Since young stellar populations can penetrate higher column densities than older populations, the escape fraction in these haloes is on average higher. Moreover, we find that around $70 \%$ of the haloes with $\Mvir < 10^8 \, \Msun$ that contain a stellar population younger than $5 \, \Myr$ have $\fEsc > 0.01$. This is the reason why low-mass haloes are thought to be the main contributors to reionization \citep[Paper I;][]{2014MNRAS.442.2560W}. Although supernova feedback is essential for ionizing radiation to escape from massive haloes, it does not play an important role in the majority of reionization sources.  

In order to make the radiative transfer simulations as detailed as possible (e.g. including helium and multi-frequency transport), our simulations were performed in post-processing. This has some drawbacks. Since the feedback from the ionizing radiation on the gas is not taken into account, we may underestimate the escape fraction in low-mass haloes in which the ionization front can evacuate gas from the halo. We found that the uniform UV background had a profound impact on the escape fraction and on the halo mass that was still able to form stars in a reionized inter-galactic medium. Following the ionizing radiation from the sources self-consistently would give a more accurate picture of the redshift evolution of the escape fraction. In addition, the ionization state of the gas inside the halo in our initial conditions does not take into account the effect of previous generations of stars, we could therefore underestimate the escape fraction in more massive haloes that form stars continuously. 

Our result that the majority of ionizing photons is absorbed by the birth cloud shows the importance of properly resolving the inter-stellar medium and the dense clouds from which the stars form. On the other hand, due to the large scatter in escape fractions, it is impossible to draw conclusions from very detailed simulations that follow only a few objects. Future studies should therefore not only strive for the highest possible resolution, but also for a large enough sample to account for the physical scatter in the escape fraction. 

One major uncertainty in this study is the output of ionizing photons from the star particles. Depending on the physics included in the stellar synthesis models, this could vary by a factor of $\sim 2 - 5$ \citep{Eldridge:2009tw,2014ApJS..212...14L}. Whether the ionization front can break out of the low-mass haloes before supernova feedback kicks in depends critically on this value, as the Str\"{o}mgren radius will vary accordingly by a factor $\sim 1.25 - 1.7$.

\subsection{Implications}

Observational studies at $z \sim 3$, the highest redshift from which ionizing photons can be detected, find evidence for escaping ionizing radiation in only a few of the observed objects \citep{2013ApJ...765...47N, Mostardi:2013ff}. Our results show that even in galaxies with very high escape fractions there can be many sight lines through which no ionizing radiation escapes. Whether any evidence for escaping ionizing radiation is observed thus highly depends on the orientation of the galaxy. This stresses the importance of large samples in observational studies. The very low mass we find for the haloes with the highest escape fractions means that these are extremely hard to observe. In order to use observations to learn more about how ionizing radiation escapes from a galaxy it is therefore necessary to look for local analogues of the high-redshift galaxies with high escape fractions. In the high-mass range we find that the escape fraction is very high in a few haloes in which feedback was so violent that it removed most of the dense gas in the centre of the halo, causing large outflows. Recent observations have indeed found evidence for leaking ionizing radiation in one such object \citep{2011ApJ...730....5H,2014Sci...346..216B}. Observationally constraining the escape fraction during the epoch of reionization may be possible by targetting strongly lensed galaxies and look for features in the spectra that indicate a high escape fraction \citep{2013ApJ...777...39Z}. This provides an interesting way to test our escape fraction predictions.

The topology of reionization has to be studied using large-volume simulations to make sure the results are unbiased \citep{Iliev:2013vl}. For this reason, numerical and semi-numerical approaches do not take into account baryonic physics but assume sources form in dark-matter haloes. The ionizing emissivity of the sources is generally taken to scale with the mass of the halo \citep[e.g.][]{Trac:2007gg,Mesinger:2007hl,2012MNRAS.423.2222I}. At first sight this seems a reasonable guess because the star formation rate is found to scale with halo mass, so haloes with a higher mass produce more ionizing photons. However, the strong mass dependence of the escape fraction that we find shows that this approach is strongly biased. Instead of by relatively few massive sources, most ionizing photons are emitted from low-mass haloes below the resolution limit of current reionization simulations. This could have a strong impact on the topology of reionization. In addition, the large range of escape fractions we find in haloes of comparable mass shows that it is unrealistic to use a single value for the escape fraction, an approach adopted by most semi-analytic, semi-numerical and numerical studies of reionization. Lastly, the strong anisotropy of the escape fraction could strongly affect the reionization topology and should therefore be taken into account as well. Upcoming observations of the distribution of neutral hydrogen during reionization provides an indirect way to test the escape fractions and reionization sources that are favoured by our simulations \citep[see e.g.][for a review]{2012RPPh...75h6901P}.


\section{Conclusions}

We have computed the escape fraction of ionizing photons in a very large sample of proto-galaxies that form during the epoch of reionization. The proto-galaxies were extracted from two cosmological hydrodynamics simulations of galaxy formation that are consistent with current observational constraints on the stellar mass function and star formation rate at high redshift (Khochfar et al in prep). Multi-frequency radiative transfer of ionizing photons was performed in post-processing. The very large sample ($75801$ haloes) combined with the high resolution ($M_{\textnormal{SPH}} = 1250 \, \Msun$) and the detailed radiative transfer simulations allows us to study the statistical properties of the escape fraction in large detail. Our main conclusions can be summarised as follows:

\begin{itemize}
    \item The column density within $10 \, \pc$ around the sources is the main constraint on the escape fraction, as most of the ionizing photons produced by young, massive stars are absorbed here. 
    \item In haloes of $\Mvir \lesssim 10^8 \, \Msun$ that have formed stars in the last $5 \, \Myr$, the ionizing radiation from the young stars can penetrate the dense gas in their surroundings, resulting in $\fEsc \gtrsim 0.1$ on average.
    \item Haloes in this mass range that have not recently formed stars have escape fractions close to zero, unless feedback has re-arranged the dense gas surrounding the stars.
    \item In haloes with $\Mvir \gtrsim 10^8 \, \Msun$, in which star formation is continuous, the larger potential well causes higher densities in the centres of these haloes ($\NH \gtrsim 2 \times 10^{22} \, \cmmt$). In $> 75\%$ of the haloes in this mass range most of the ionizing radiation from young stars is absorbed by the birth cloud, resulting in an escape fraction less than $1\%$.
    \item Haloes of $\Mvir \gtrsim 10^8 \, \Msun$ have $\fEsc > 0.01$ only if supernova feedback either causes a inhomogeneous density around the young sources ($\sim 70 \%$ of the cases) or clears away all the dense gas around some of the young sources ($\sim 30 \%$ of the cases).
    \item There is a large scatter in escape fractions whose origin is physical. Principal component analysis shows that using only one or two variables to parametrize the escape fraction leads to biased results. 
    \item The probability density function of the escape fraction in different mass bins shows that the distribution is peaked towards high escape fraction ($\fEsc \gtrsim 0.1$) for haloes in which $\Mvir < 10^8 \, \Msun$, while shifting  towards much smaller escape fractions at higher masses.
    \item The uniform UV-background has a profound impact on the escape fraction and ionizing photon production rate in haloes with $\Mvir \lesssim 5 \times 10^8 \, \Msun$. Photo-evaporation leads to very high escape fractions in this mass range, but star formation is suppressed, so these haloes no longer contribute to cosmic reionization. 
    \item $\HeI$-ionizing photons escape more easily than $\HI$-ionizing photons, because the cross section at this frequency is lower. Almost no $\HeII$-ionizing photons escape because most are absorbed in the halo itself.
    \item The escape of ionizing radiation is highly anisotropic. In haloes hosting stellar populations younger than $5 \, \Myr$ the average solid angle through which radiation escapes is $\SA \sim 2 \pi \, \sr$ for $\fEsc > 0.5$ and drops rapidly to $\SA \sim 0.3 \pi \, \sr$ for $\fEsc = 0.01$.
    \item Ionising radiation escapes primarily through one or two sight lines. A larger $\SA$ means that the patches through which radiation escapes are larger, not that there are more.
\end{itemize}

Our results stress the important role of proto-galaxies forming in haloes of $\Mvir \lesssim 10^8 \, \Msun$ for cosmic reionization. These haloes are more abundant and have higher escape fractions than their higher mass counterparts and are therefore essential to take into account in models of reionization. The large range of escape fractions we find in haloes with similar properties shows that to study the escape fraction either observationally or numerically, large samples are essential. Numerical simulations of the topology of reionization on large scales could greatly improve by taking this scatter into account. 

In this work we have focussed on the escape fraction in the entire sample of haloes and the redshift evolution of the average escape fraction. In individual haloes the redshift evolution of the escape fraction can deviate a great deal from this average. We plan to study this in greater detail in future work.

\section*{Acknowledgments}
The authors would like to thank Eyal Neistein for generating the merger tree, Daniel Schaerer for providing the Pop III spectra, Chael Kruip for providing the basis of the ray-tracing routine, and the anonymous referee for comments that helped to improve the clarity of the presentation. These simulations were run using the facilities of the Rechenzentrum Garching. JPP acknowledges support from the European Research Council under the European Communitys Seventh Framework Programme (FP7/2007-2013) via the ERC Advanced Grant "STARLIGHT: Formation of the First Stars" (project number 339177). CDV acknowledges support by Marie Curie Reintegration Grant PERG06-GA-2009-256573. This research made use of Astropy, a community-developed core Python package for Astronomy \citep{2013A&A...558A..33A}.

\bibliographystyle{../include/mn2e} 

\appendix

\section[]{The effect of different stellar synthesis models for Pop II stars}\label{sec:appendix_s99vsbc03}

In this work we use the Starburst 99 package \citep{Leitherer:1999jt} to compute the spectra of the Pop II sources in the simulation, to be consistent with subsequent studies of the FiBY simulations. Here we show the difference in using the ionizing emissivity of Starburst 99 and the models of \citet{2003MNRAS.344.1000B} that were used in Paper I.

\begin{figure} 
  \includegraphics[width=85mm]{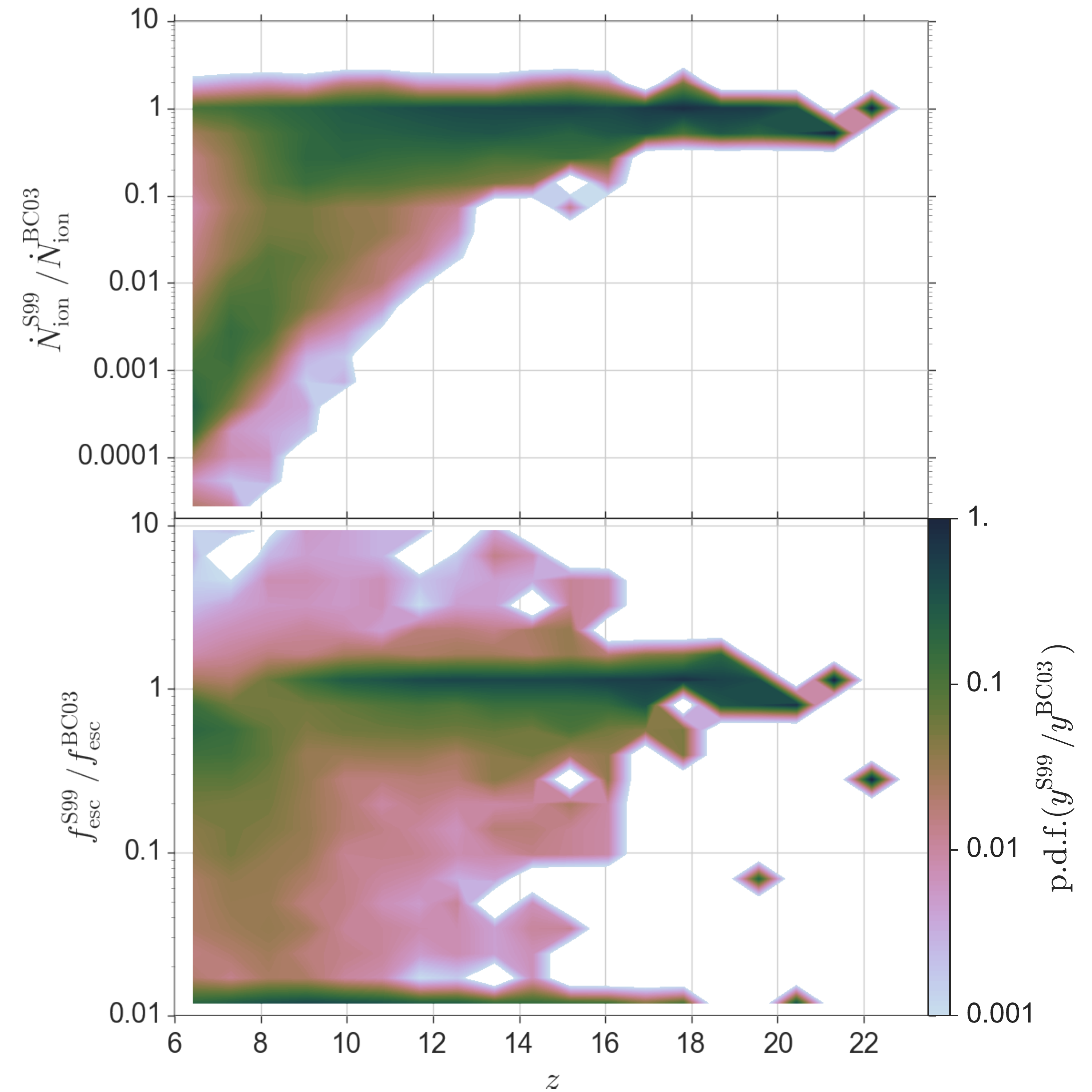}
  \caption{
{\it Top panel:} The ratio of the number of ionizing photons computed with Starburst 99 and \citet{2003MNRAS.344.1000B} in every halo in the FiBY\_S simulation as a function of redshift. {\it Bottom panel:} The ratio of the escape fraction computed with Starburst 99 and \citet{2003MNRAS.344.1000B} in every halo in the FiBY\_S simulation as a function of redshift.
  }
  \label{fig:S99_vs_BC03}
\end{figure}

In the top panel of \autoref{fig:S99_vs_BC03} we compare the ionizing photon production in all haloes in the FIBY\_S simulation for the two stellar synthesis models. At redshifts higher than $13$ the difference is small, most haloes have a comparable total ionizing emissivity. The scatter is mainly caused by differences in the ionizing photon emissivity as a function of the metallicity computed by the stellar synthesis models. This changes at lower redshifts, because the \citet{2003MNRAS.344.1000B} models give a higher ionizing photon output for stellar populations older than $100 \, \Myr$. Because the ionizing emissivity of the older stellar populations is still almost 5 orders of magnitude below the emissivity of stellar populations younger than $5 \, \Myr$, this mainly affects low-mass haloes that host no young stellar populations. Due to quenching of star formation in the (abundant) low-mass haloes by the UV-background between redshift $12$ and $6$, we see the largest differences in ionizing emissivity there.

In the bottom panel of \autoref{fig:S99_vs_BC03} we show the difference in escape fraction computed using Starburst99 and \citet{2003MNRAS.344.1000B} in all haloes in the FiBY\_S simulation. At redshifts higher than $13$ the difference in escape fractions is small, as the difference in ionizing emissivity is small as well. Between redshift $13$ and $8$ the scatter grows because there are more and more old stellar populations, but the majority of haloes have comparable escape fractions. This changes only between redshift $8$ and $6$, where the largest differences in ionizing emissivity occur which reflect on the escape fractions. At these redshifts the difference in escape fractions in the majority of haloes is around a factor 2. 

On average the escape fraction computed with the BC03 models is higher than those computed with S99. This is expected if the total number of absorptions in the halo is the same independent of the source model used. This is the case if the gas around the old populations is highly ionized and most of the photons produced in these sources reach the virial radius. However, in some cases the old populations are obscured, which results in more absorptions and thus a lower escape fraction in case the BC03 models are used.

These results show not only the importance of stellar synthesis models for computing escape fractions, but also the potential importance of older stellar populations. These populations are often neglected because their ionizing photon output is so low, but if there are many old stellar populations in a halo they could still have an impact, especially because the column density around these sources is likely lowered by supernova feedback. However, our simulations indicate that the contribution from old sources only starts kicking in after reionization is completed and thus S99 or BC03 will provide consistent results.

\section[]{The difference in escape fraction between continuous and bursty star formation}\label{sec:appendix_constVsBurstSF}

\begin{figure} 
  \includegraphics[width=85mm]{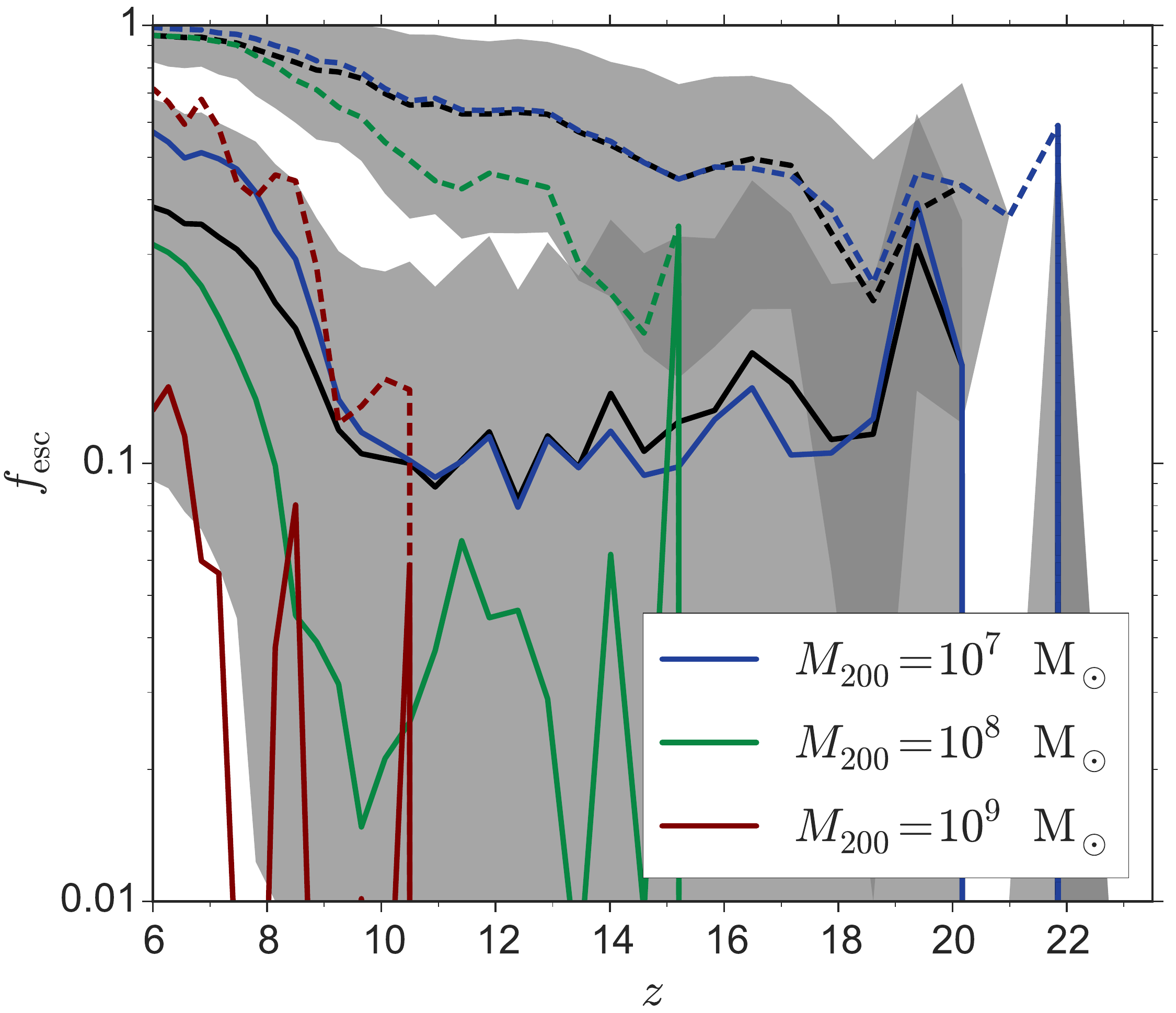}
  \caption{The average escape fraction as a function of redshift for models of burst-like star formation (solid lines) and continuous star formation (dashed lines) for the FiBY\_S simulation. Black lines represent the mean of all haloes, with the shaded area the standard deviation. The coloured lines represent the average escape fraction in mass bins of $10^7 \, \Msun$ (blue), $10^8 \, \Msun$ (green) and $10^9 \, \Msun$ (red).}
  \label{fig:BC03_burst_vs_cont}
\end{figure}

In Paper I we assumed that star formation in each star particle was continuous, which results in an strict upper limit of the escape frac- tion. In \autoref{fig:BC03_burst_vs_cont} we show the average escape fraction as a function of redshift for the \citet{2003MNRAS.344.1000B} models assuming bursty (solid lines) and continuous (dashed lines) star formation. The difference in average escape fraction is largest around redshift 10, with an average escape fraction of $\sim 0.1$ for bursty star formation and $\sim 0.6$ for continuous star formation. After the UV-background comes into effect the differences become smaller. 

The average escape fraction is dominated by the most-abundant low-mass haloes. In these haloes the difference in average escape fraction (blue lines) is mainly caused by the population of haloes that have not formed stars in the last $5 \, \Myr$. In the bursty star formation runs this resulted in a population of haloes that have very low escape fractions, a population that doesn't exist in the continuous star formation runs. The escape fractions in low-mass haloes that have recently formed stars are comparable. In haloes of higher mass (green and red lines) the differences in escape fractions are larger because these haloes contain multiple generations of stars whose ionizing emissivity adds up in the continous star formation runs. This strengthens our conclusions in Paper I that the lowest-mass haloes contribute most ionizing photons to reionization because we have overestimated the ionizing emissivity most severly in the more massive haloes. Although the difference between the two runs is considerable, it is not much larger than the difference between the two stellar synthesis models discussed in \autoref{sec:appendix_s99vsbc03} and the general trends we observed in Paper I remain valid.

\bsp

\label{lastpage}

\end{document}